\documentclass[review,authoryear,3p,times,12pt]{elsarticle}

\usepackage{multirow,setspace,times,amssymb,amsmath,graphicx,color,rotating,subfigure,url}
\usepackage{lineno,color}
\usepackage{natbib}
\usepackage{booktabs}%
\usepackage{longtable}%
\usepackage{rotating}
\usepackage{lscape}
\usepackage{threeparttable}
\graphicspath{{Figures/}}
\usepackage[table]{xcolor}
\usepackage{tabularx}
\usepackage{graphicx} 
\usepackage{epstopdf}
\usepackage{mathrsfs}
\usepackage{makecell}
\usepackage[bookmarks=true,colorlinks,linkcolor=blue,anchorcolor=blue,citecolor=blue,unicode,urlcolor=blue]{hyperref}
\usepackage{bookmark}
\usepackage{todonotes}
\usepackage{ragged2e}
\usepackage[font=small]{caption}

\usepackage{lineno}
\usepackage{dcolumn}
\hypersetup{CJKbookmarks=true}%

\begin{document}
\captionsetup[figure]{font={bf},name={Fig.},labelsep=period}
\captionsetup[table]{labelfont={bf},name={Table},labelsep=newline,singlelinecheck=false}

\begin{frontmatter}

\title{Uncertainty and financial market resilience: Evidence from China}	
\author[SB,RCE]{Si-Yao Wei}
\author[SEM]{Kun-Liang Jiang}
\author[SB,RCE,Math]{Wei-Xing Zhou\corref{cor1}}
\ead{wxzhou@ecust.edu.cn} 
\cortext[cor1]{Corresponding author.}
\address[SB]{School of Business, East China University of Science and Technology, Shanghai 200237, China}
\address[RCE]{Research Center for Econophysics, East China University of Science and Technology, Shanghai 200237, China}
\address[SEM]{School of Economics and Management, Southwest University of Science and Technology, Mianyang 621010, China}
\address[Math]{School of Mathematics, East China University of Science and Technology, Shanghai 200237, China}

\begin{abstract}
Financial market resilience reflects the ability of a financial market to withstand external shocks and to recover from them, while its measurement has yet to be standardized. Accordingly, this paper quantifies the adaptability and recoverability of China's total financial market and five key sub-markets as both proxy indicators of their resilience. The results highlight the event-driven nature of China's financial market resilience and reveal a strong correlation between the two indicators, which is more pronounced in the stock and bond markets. Using the Diebold-Yilmaz connectedness approach, we further examine volatility spillovers among resilience of sub-markets and identify the foreign exchange market as a major transmitter of spillovers, whereas the stock and bulk commodity markets primarily act as net recipients of spillovers. Moreover, we analyze the impacts of five China-related uncertainties on financial market resilience. Overall, geopolitical risks, economic and trade policy uncertainty, and U.S.-China tensions exert significant negative impacts on total market resilience, while the effect of climate policy uncertainty remains less pronounced. Importantly, the impacts of different uncertainties exhibit heterogeneity across resilience of sub-markets. Our findings not only enrich the resilience measurement of financial market but also provide new evidence to inform targeted risk management and policy design aimed at strengthening financial system's resilience.
\end{abstract}

\end{frontmatter}

\section{Introduction}

Since the outbreak of the 2008 global financial crisis, the vulnerability of financial systems has been further magnified, and the characteristics of risk propagation have become increasingly complex. Financial crises nowadays have shown that adverse developments in one market can have dramatic effects on others, even when the latter possess seemingly strong fundamentals and distinct structures \citep{FJ-Cevik-Terzioglu-Kilic-Bugan-Dibooglu-2024-ResIntBusFinanc}. Meanwhile, escalating geopolitical risks and global economic downturns have made financial markets more unstable \citep{FJ-Nguyenhuu-Orsal-2024-WorldEcon}. For example, during the Russia-Ukraine conflict, the international foreign exchange and bulk commodity markets experienced significant volatility due to restrictions on Russia's use of the U.S. dollar, the Euro, and the Japanese yen for international transactions. In this context of heightened uncertainty, clarifying how to maintain the stability of financial systems and markets has become increasingly important, as emphasized in prior literature \citep{FJ-Battiston-Gatti-Gallegati-Greenwald-Stiglitz-2012-JEconDynControl,FJ-Blot-Creel-Hubert-Labondance-Saraceno-2015-JFinancStab,FJ-GreenwoodNimmo-Tarassow-2016-EconModel}.

Specifically, there have been studies that investigate the impact of geopolitical risks on financial stability from multiple perspectives \citep{FJ-Nguyenhuu-Orsal-2024-WorldEcon,FJ-Zhu-Xia-Li-Chen-2025-FinancResLett}, while other scholars have examined whether climate-related uncertainty affects financial market volatility \citep{FJ-Battiston-Dafermos-Monasterolo-2021-JFinancStab,FJ-Liang-Goodell-Li-2024-JIntFinancMarkInstMoney}. Furthermore, extensive studies have analyzed the relationship between economic policy uncertainty and volatility of financial market \citep{FJ-You-Guo-Zhu-Tang-2017-EnergyEcon,FJ-Wang-Li-He-2020-ResIntBusFinanc,FJ-Phan-Iyke-Sharma-Affandi-2021-EconModel}. Overall, it is well established that monitoring financial market stability is closely related to controlling market volatility \citep{FJ-Mamatzakis-Tsionas-2021-IntJFinancEcon}, especially during disruptions, offering valuable insights into the mechanisms underlying financial stability.

However, on the one hand, as market interconnectivity deepens, merely controlling volatility in a single market or implementing short-term stabilization measures cannot eliminate shocks transmitted through inter-market linkages \citep{FJ-Mieg-2022-RiskAnal,FJ-Kubitza-2025-JFinancQuantAnal}. On the other hand, stability typically concerns equilibrium and minor deviations, it fails to characterize large-scale system transitions, adaptive governance, and long-term recovery processes \citep{FJ-Folke-2006-GlobEnvironChange,FJ-Haldane-May-2011-Nature}. This suggests that focusing solely on financial market stability may be insufficient to capture true risk contagion \citep{FJ-Forbes-Rigobon-2002-JFinanc}. Consequently, there is broad consensus on the need to examine what the resilience of a financial system or market is, and how to build a resilient financial system or market \citep{FJ-Bui-Scheule-Wu-2017-JFinancStab,FJ-Koetter-Krause-Sfrappini-Tonzer-2022-EurEconRev}.

The concept of resilience originates in ecology, where \cite{FJ-Holling-1973-AnnuRevEcolSystemat} defined it as a system's capacity to withstand disturbances without collapsing or transitioning to a fundamentally altered state, and to recover its essential functions afterward. Building on this foundation, \cite{FJ-Bruneau-Chang-Eguchi-Lee-ORourke-Reinhorn-Shinozuka-Tierney-Wallace-VonWinterfeldt-2003-EarthqSpectra} proposed a deterministic metric to measure a community's performance loss during an earthquake, which is regarded as that community's resilience \citep{FJ-Hosseini-Barker-RamirezMarquez-2016-ReliabEngSystSaf}. Subsequently, within this framework\textemdash known as the {\it Resilience Triangle}\textemdash various disciplines have developed their own definitions and measurements of resilience. For instance, engineering studies often assess resilience by measuring the cumulative loss in infrastructure quality over time \citep{FJ-Sahebjamnia-Torabi-Mansouri-2015-EurJOperRes,FJ-Chen-Ma-Chen-Yang-2024-TranspResPartD}, while management research defines resilience through changes in supply chain performance indicators \citep{FJ-Ivanov-2022-AnnOperRes} or through simulations of disruption impacts on supply systems \citep{FJ-Ghadge-Er-Ivanov-Chaudhuri-2022-IntJProdRes}. In ecology, resilience is typically quantified by the recovery rate or the time required to adapt to changing disturbance regimes \citep{FJ-Seidl-Spies-Peterson-Stephens-Hicke-2016-JApplEcol}.

In the finance field, resilience is primarily understood as the ability of agents\textemdash such as individuals, firms, and financial institutions\textemdash to withstand external shocks and to recover from them \citep{FJ-Haimes-2009-RiskAnal,FJ-Bui-Scheule-Wu-2017-JFinancStab,FJ-Chen-Sun-Zhang-2025-JFinancStab}. Some scholars conceptualize financial resilience in terms of risk levels \citep{FJ-SakyiNyarko-Ahmad-Green-2022-JDevStud,FJ-Daadmehr-2024-RiskManag}. For instance, \cite{FJ-Ciullo-Strobl-Meiler-Martius-Bresch-2023-NatCommun} developed a risk diversification metric based on Value-at-Risk (VaR) to examine how global catastrophe risk pooling can enhance countries' financial resilience, while \cite{FJ-Bui-Scheule-Wu-2017-JFinancStab} analyzed unconditional losses to the financial system to explore the relationship between bank capital buffers and the resilience of the Australian financial system. Although these approaches capture the risk-resistant dimension of resilience, they tend to neglect the recovery capacity during and after shocks, and thus only partially reflect the true meaning of resilience. Other studies have attempted to measure financial resilience through comprehensive indicator systems \citep{FJ-Chen-He-2022-Sustainability,FJ-Xu-Liu-2024-EmergMarkRev}. However, such methods are often subject to biases in indicator selection, which may compromise the reliability of the results. Therefore, developing richer and more objective measures of financial resilience is essential for advancing research on how to sustain a resilient financial system and market.

Given that the {\it Resilience Triangle} framework provides an integrated perspective encompassing recovery time, vulnerability, and resilience loss \citep{FJ-Klimek-Poledna-Thurner-2019-NatCommun,FJ-Fang-Chu-Fu-Fang-2022-TranspResPartD}, it is regarded as one of the most representative approaches for measuring financial resilience. Building upon this framework, \cite{FJ-Liu-Zhang-Li-2021-cnSSC} introduced the concepts of absorption intensity and duration to quantify China's macroeconomic resilience during systemic risk shocks. Subsequently, \cite{FJ-Tang-Liu-Zhou-2022-JIntFinancMarkInstMoney} measured global financial market resilience against short-term capital flow shocks, while \cite{FJ-Tang-Liu-Yang-2024-Pac-BasinFinancJ} assessed the resilience of China's financial markets to hot money shocks. Further applications can also be found in \cite{FJ-Chen-Sun-2024-EconLett} and \cite{FJ-Chen-Sun-Zhang-2025-JFinancStab}. However, compared with the extensive literature on financial stability, research on the measurement of financial market resilience remains relatively limited and warrants further exploration. In addition, existing studies primarily focus on measuring the resilience of the overall market, while the interconnections among the resilience of different sub-markets\textemdash except in rare studies like \cite{FJ-Tang-Liu-Yang-2024-Pac-BasinFinancJ}\textemdash are often neglected, which may result in an incomplete understanding of a system's dynamics \citep{FJ-Wang-Xie-Jiang-Stanley-2016-FinancResLett}. Furthermore, although recent studies have begun to explore the relationship between uncertainty and financial market resilience, such as climate-related uncertainty \citep{FJ-Zhang-Shi-Zhao-Yang-2025-RiskManag,FJ-Yao-Maimaitijiang-Li-Le-2025-JCommodMark}, comprehensive comparisons of how various types of uncertainty heterogeneously affect resilience remain underexplored.

Consequently, using China's financial market as a case study, our research aims to address these limitations. First, within the {\it Resilience Triangle} framework, we quantify the adaptability and recoverability of the financial market under external disruptions as its resilience. To characterize the state pathway of a market during shocks, we employ the time-varying parameter vector autoregressive (TVP-VAR) model to obtain the impulse response function, which effectively captures the magnitude, persistence, and transmission effects of shocks \citep{FJ-Klimek-Poledna-Thurner-2019-NatCommun,FJ-Brunnermeier-2024-JFinanc}. Second, given the strong interdependence in volatility across markets, we utilize the TVP-VAR-DY connectedness model to examine the dynamic linkages among the resilience of five financial sub-markets. This approach allows us to identify how changes in one market's resilience propagate to others. Third, we introduce five uncertainty indices to investigate their heterogeneous impacts on financial market resilience. Specifically, we analyze both the average effects of uncertainty on resilience using a baseline regression model and the dynamic spillover effects through the TVP-VAR-DY framework. Overall, our findings provide new evidence on the evolution of financial resilience in China and offer insights into monitoring how changes in external uncertainties affect systemic stability.

This paper makes two main contributions to the literature on uncertainty and financial market resilience. First, it enriches the measurement of financial market resilience by adopting a two-dimensional framework that simultaneously quantifies a market's ability to adapt and recover from shocks. Compared with a single-dimension measure such as risk resistance, this approach offers a more comprehensive depiction of the essence of resilience. Moreover, to ensure consistent directional interpretation, our definition of recoverability differs from the duration measure used in previous studies \citep{FJ-Tang-Liu-Zhou-2022-JIntFinancMarkInstMoney,FJ-Chen-Sun-2024-EconLett}. Second, this paper advances the literature on uncertainty and resilience by revealing how China-related uncertainties influence its financial market resilience to external shocks. Compared with global uncertainty indicators, using China-specific uncertainties helps simplify the channels of risk transmission and clarify their mechanisms. Furthermore, beyond static regression analysis, our examination of time-varying impacts captures the dynamic characteristics of risk spillovers\textemdash particularly during major events\textemdash thereby providing valuable implications for regulatory authorities in managing financial risks.

The subsequent sections of this paper are structured in the following manner. The approach is explicated in Section~\ref{Sec_Methodology}, and Sections~\ref{Sec_Data} and~\ref{Sec_Empirical} offer and examine the empirical data and primary discoveries. In conclusion, Section~\ref{Sec_Conclusion} serves as the final segment of the paper.

\section{Methodology}
\label{Sec_Methodology}

\subsection{Measuring the financial market resilience}

Adaptability (or reliability) and recoverability are considered two essential principles of resilience \citep{FJ-Haimes-2009-RiskAnal,FJ-Hosseini-Barker-RamirezMarquez-2016-ReliabEngSystSaf}. A greater adaptability or recoverability represents a better resilience. Hence, according to \cite{FJ-Liu-Zhang-Li-2021-cnSSC}, we utilize the {\it Resilience Triangle} framework to measure both indicators as China's financial market resilience \citep{FJ-Tang-Liu-Zhou-2022-JIntFinancMarkInstMoney,FJ-Brunnermeier-2024-JFinanc}.

\begin{figure}[htb!]
\centering
\includegraphics[width=0.8\textwidth]{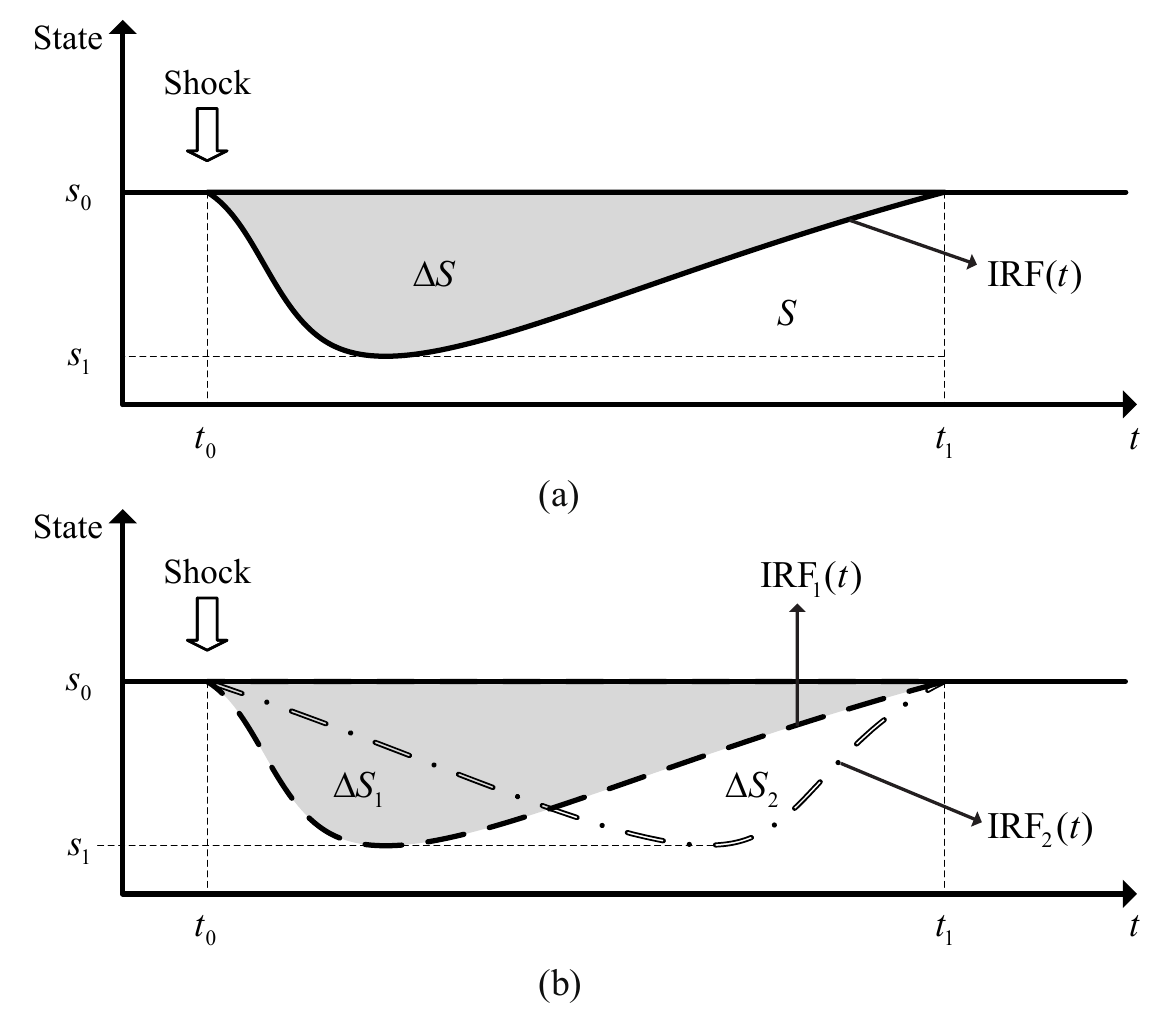}
\caption{The state pathway of the financial market to the shock. \normalfont{In this figure, $\mathrm{IRF}(t)$ refers to the impulse response function. $s_0$ and $s_1$ refer to the original state and the state at which the shock strength is maximized, respectively. $\Delta S$, $\Delta S_1$, and $\Delta S_2$ all represent the areas of the regions enclosed by the impulse response functions and the original level line, while $S$ represent the blank area.}}
\label{Fig_Res_Definition}
\end{figure}

Generally, the state of a system varies when facing an external shock. Afterward, its subsequent evolution may follow three possible pathways \citep{FJ-Brunnermeier-2024-JFinanc}: (1) a positive pathway, where the system recovers to a state better than its original level; (2) a negative pathway, where it recovers to a worse state; (3) an original pathway, where it returns to its original state. For simplicity, we focus on the original pathway and introduce the impulse response function (IRF) to depict the evolution of its pathway in this paper. As shown in Fig.~\ref{Fig_Res_Definition}a, when a financial market faces an external shock at $t_0$, its state varies from $s_0$ to $s_1$ and recovers to the original state at $t_1$. Let us consider changes in both state and recovery time, the $\mathrm{IRF}(t)$ splits the original state into two parts: $\Delta S$ and $S$. The larger $S$, the less the market is affected by the shock. Hence, the adaptability of the market $i$ conditional on the shock $j$ at time $t$ can be expressed as:
\begin{equation}
\label{eq_adaptability}
    R^\mathrm{Adapt}_{t,i\leftarrow j}
    =\frac{S}{S+\Delta S}
    =\frac{N\times \max \left\{ \left| \Phi _{t,i\leftarrow j}^{n} \right| \right\}-\sum\limits_{n=1}^{N} D_{t,i\leftarrow j}^{n}}{N\times \max \left\{ \left| \Phi _{t,i\leftarrow j}^{n} \right| \right\} },
\end{equation}
where $N=t_1-t_0$ denotes the entire period during the shock. $\Phi_{t,i\leftarrow j}^{n}$ denotes the IRF value of the financial market $i$ to the shock $j$ at the $n$-th period. $D_{t,i\leftarrow j}^{n} = \max \left\{ \left| \Phi _{t,i\leftarrow j}^{n} \right| \right\}-\left| \Phi _{t,i\leftarrow j}^{n} \right|$ measures the performance gap between the market's current state and its state of maximum disruption, capturing the extent of recovery at the $n$-th period relative to the worst-case scenario. Notably, we do not differentiate the direction in which the shock causes the market to change.

Furthermore, Fig.~\ref{Fig_Res_Definition}b exhibits that $\mathrm{IRF}_1(t)$ and $\mathrm{IRF}_2(t)$ represent different pathways but there may exist $\Delta S_1=\Delta S_2$, which indicates that Eq.~(\ref{eq_adaptability}) is insufficient for determining their resilience. Hence, we introduce the concept of ``center of gravity'', which suggests that the pathway of $\mathrm{IRF}_1(t)$ is more resilient than that of $\mathrm{IRF}_2(t)$ due to the less delayed ``center of gravity'' and the earlier recovery \citep{FJ-Tang-Liu-Yang-2024-Pac-BasinFinancJ,FJ-Chen-Sun-Zhang-2025-JFinancStab}. Subsequently, the recoverability of the market $i$ conditional on the shock $j$ at time $t$ can be expressed as:
\begin{equation}
\label{eq_recoverability}
    R^\mathrm{Recover}_{t,i\leftarrow j}
    =\sum\limits_{n=1}^{N}{n}\times w_{t,i\leftarrow j}^{n}
    =\frac{\sum\limits_{n=1}^{N}{n} \times D_{t,i\leftarrow j}^{n}}{\sum\limits_{n=1}^{N} D_{t,i\leftarrow j}^{n}},
\end{equation}
where the dynamic weight $w_{t,i\leftarrow j}^{n}={D_{t,i\leftarrow j}^{n}}/{\sum\limits_{n=1}^{N}D_{t,i\leftarrow j}^{n}}$. This formula essentially calculates a time-weighted average of the performance gaps, where periods with larger gaps (slower recovery) receive more weight. A lower value of $R^\mathrm{Recover}_{t,i\leftarrow j}$ thus indicates a slower, more delayed recovery process, akin to a center of gravity further away from the initial shock.

Overall, a higher $R^\mathrm{Adapt}_{t,i\leftarrow j}$ indicates a stronger ability to absorb risks and thus greater adaptability of the financial market. Similarly, a higher $R^\mathrm{Recover}_{t,i\leftarrow j}$ indicates a faster recovery toward the original level and hence greater recoverability. Notably, for easier comparison between the two indicators, the dynamic weight used in Eq.~(\ref{eq_recoverability}) differs from that in some previous studies \citep{FJ-Tang-Liu-Zhou-2022-JIntFinancMarkInstMoney,FJ-Tang-Liu-Yang-2024-Pac-BasinFinancJ,FJ-Chen-Sun-2024-EconLett}. These scholars emphasize absorption duration to shocks, where the shorter duration denotes the better resilience. Yet, we focus more on the recoverability of a market, where the greater recoverability denotes the better resilience, thereby ensuring consistent directional interpretation of both indicators.

\subsection{Offering impulse response function: The TVP-VAR model}

To obtain the time-varying IRFs necessary for our dynamic resilience measures, we employ the TVP-VAR model \citep{FJ-Primiceri-2005-RevEconStud}. Unlike standard VAR models, the TVP-VAR framework allows for parameter drift and stochastic volatility, which is crucial for capturing the evolving structure of China's financial markets over the long sample period. The model is defined as follows:
\begin{equation}
\label{eq_tvpvar}
    y_t=c_t+B_{1,t}y_{t-1}+\cdots +B_{s,t}y_{t-s}+e_t,\quad e_t\sim N(0,\Omega_t),
\end{equation}
where $t=s+1,\ldots, n$, $y_t$ is the $k\times1$ dimensional observation vector, $c_t$ is an $k\times 1$ vector of time-varying coefficients that multiply constant terms, $B_{1,t},\ldots,B_{s,t}$ is the time-varying coefficients matrix of $k\times k$, and $e_t$ is heteroscedastic unobservable shock with $\Omega_t$ \citep{FJ-Chen-Sun-Zhang-2025-JFinancStab}. $\Omega_t$ is the time-varying covariance matrix of $k\times k$, which is decomposed into 
$\Omega_t = A_t^{-1} \Sigma_t \Sigma_t A{'}_t^{-1}$, where $A_t$ is a lower triangular matrix with diagonal equal to 1 and $\Sigma_t={\mathrm{diag}}(\sigma_{1,t},\ldots ,\sigma_{k,t})$. Suppose that $\alpha_t$ and $\beta_t$ are the row vectors of the matrices $A_t$ and $B_{1,t},\ldots, B_{s,t}$, respectively, and that $h_t=(h_{1,t},\ldots,h_{k,t})$, where $h_{i,t}=\log \sigma_{i,t}^2$. The time-varying parameters obey the following stochastic process:
\begin{align}
\label{eq_stochastic}
    \begin{matrix}
   \beta_{t+1}=\beta_t+u_{\beta,t},  \\
   \alpha_{t+1}=\alpha_t+u_{\alpha,t},  \\
   h_{t+1}=h_t+u_{h,t},  \\
\end{matrix}\quad \left(\begin{matrix}
   \varepsilon_t  \\
   u_{\beta,t}  \\
   u_{\alpha,t}  \\
   u_{h,t}  \\
\end{matrix} \right)\sim N\left(0,\left( \begin{matrix}
   I & O & O & O  \\
   O & \Sigma_\beta & O & O  \\
   O & O & \Sigma_\alpha & O  \\
   O & O & O & \Sigma_h  \\
\end{matrix} \right) \right)
\end{align}
where $\varepsilon_t=\Sigma _t^{-1}A_t e_t$. The time-varying parameter vectors satisfy their respective conditional distributions: $\beta_{s+1}\sim N(\mu_{\beta_0}, \Sigma_{\beta_0})$, $\alpha_{s+1}\sim N(\mu_{\alpha_0}, \Sigma_{\alpha_0})$, and $h_{s+1}\sim N(\mu_{h_0},\Sigma_{h_0})$. Based on Eqs.~(\ref{eq_tvpvar}) and (\ref{eq_stochastic}), we can obtain the impulse response value under the 1-unit standardized shock at each time point.

\subsection{Measuring dynamic connectedness: The TVP-VAR-DY model}

To explore how resilience volatility propagates across different sub-markets, we extend the analysis by the connectedness methodology, which can be used to measure the impact of returns or return volatilities in some market or asset on other markets or assets \citep{FJ-Diebold-Yilmaz-2012-IntJForecast}, and is usually time-varying. There are several studies on dynamic connectedness using the way combined with rolling window and DY model \citep{FJ-Tiwari-Suleman-Ullah-Shahbaz-2023-IntJFinancEcon,FJ-Yang-Geng-Liang-2024-EnergyEcon}. However, the window size setting in this method is subjective and inevitably loses some observations. To capture the dynamic changes in the underlying structure of the data more flexibly, according to \cite{FJ-Antonakakis-Chatziantoniou-Gabauer-2021-IntJFinancEcon}, we use the TVP-VAR model to characterize the changes of time-varying parameters and combine it with the DY model to measure the connectedness in different time domains.

The equations for time-varying parameters of the TVP-VAR model have been defined by Eqs.~(\ref{eq_tvpvar}) and (\ref{eq_stochastic}), and only how to combine this model with the DY model is presented here. Extending the TVP-VAR model into a TVP-VMA model:
\begin{equation}
    y_t=c_t+B_{1,t}y_{t-1}+\cdots +B_{s,t}y_{t-s}+e_t=\mu_t+\sum\limits_{h=0}^{\infty}\Lambda_{h,t}\varepsilon_{t-h},
\end{equation}
where $\Lambda_{h,t}=\sum\limits_{j=1}^{h}\Lambda_{h-j,t}B_{j,t}$ denotes the matrix of time-varying impulse response functions and indicates the influence of a shock at $t-h$ time to a variable at $t$ time. $\mu_t$ denotes the time-varying conditional mean and $\varepsilon_t$ denotes the standardize shock.

Subsequently, based on the time-varying coefficients matrix $B_{1,t},\ldots,B_{s,t}$ computed in Eq.~(\ref{eq_tvpvar}) and the covariance matrix $\Omega_t$, the generalized impulse response functions (GIRF) and the generalized forecast error variance decomposition (GFEVD) of order $H$ can be respectively computed as:
\begin{equation}
    \Psi_{j\leftarrow i,t}(H)=\frac{\sigma_{ii,t}^{-1}\sum_{h=0}^{H-1}\left(\theta_j^{\prime}\Lambda_{h,t}\Omega_t\theta_i\right)^2}{\sum_{h=0}^{H-1}\left(\theta_j^{\prime}\Lambda_{h,t}\Omega_t\Lambda_{h,t}^{\prime}\theta_j\right)}
\end{equation}
and
\begin{equation}
    \tilde{\Phi}_{j\leftarrow i,t}(H)=\frac{\Psi_{j\leftarrow i,t}(H)}{\sum_{i=1}^k\Psi_{j\leftarrow i,t}(H)},
\end{equation}
where $\sigma_{ii,t}^{-1}$ denotes the $i$-th diagonal element of $\Sigma_t$. $\theta_j$ and $\theta_i$ are $k\times 1$ selection vectors, where they respectively have 1 in the $j$-th and $i$-th positions and 0 elsewhere. Hence, the total connectedness indices ($TCI$), the directional connectedness indices from others ($FROM$), the directional connectedness indices to others ($TO$), the net connectedness indices ($NET$), and the net pairwise directional connectedness indices ($NPDC$) can be expressed as:
\begin{equation}
    TCI_t(H)=\frac{\sum_{j,i=1,j\neq i}^k\tilde{\Phi}_{j\leftarrow i,t}(H)}{\sum_{j,i=1}^k\tilde{\Phi}_{j\leftarrow i,t}(H)} \times 100,
\end{equation}
\begin{equation}
    FROM_{j\leftarrow\bullet,t}(H)=\sum\limits_{i=1,j\neq i}^{k}\tilde{\Phi}_{j\leftarrow i,t}(H)\times 100,
\end{equation}
\begin{equation}
    TO_{\bullet\leftarrow j,t}(H)=\sum\limits_{i=1,j\neq i}^{k}\tilde{\Phi}_{i\leftarrow j,t}(H)\times 100,
\end{equation}
\begin{equation}
    NET_{j,t}(H)=TO_{\bullet\leftarrow j,t}(H)-FROM_{j\leftarrow\bullet,t}(H),
\end{equation}
and
\begin{equation}
    NPDC_{j\leftarrow i,t}(H)=\left({\tilde{\Phi}_{i\leftarrow j,t}(H)}-{\tilde{\Phi}_{j\leftarrow i,t}(H)}\right) \times 100,
\end{equation}
where a higher $TCI$ indicates a higher overall connectedness of the financial system. Also, a higher $FROM_{j\leftarrow\bullet}$ or $TO_{\bullet\leftarrow j}$ indicates a higher directional connectedness of $j$ from others or that of $j$ to others. $NET_j>0$ indicates a higher directional connectedness of $j$ to others, and $NPDC_{j\leftarrow i}>0$ indicates a higher net pairwise directional connectedness from $j$ to $i$.

\subsection{Regression model of uncertainty on China's financial market resilience}

Economic and financial conditions respond significantly to increased uncertainty \citep{FJ-Glebocki-Saha-2024-JIntFinancMarkInstMoney}. To investigate whether and how China-related uncertainty impacts the resilience of China's financial market, we construct the following regression model:
\begin{equation}
    R_t^\text{Adapt/Recover}=\gamma_0+\gamma_1\text{Uncertainty}_t+\gamma_c\text{Control}_{i,t}+\varepsilon_t,
\end{equation}
where $\text{Uncertainty}_t$ denotes an uncertainty index at $t$ time, including China's geopolitical risk (CGPR) index, China's economic policy uncertainty (CEPU) index, China's climate policy uncertainty (CCPU) index, the U.S.-China tension (UCT) index, and China's trade policy uncertainty (CTPU) index. The vector $\text{Control}_{i,t}$ includes a set of standard macroeconomic variables at $t$ time to control for the domestic macroeconomic cycles including Industrial Added Value (IAV), Consumer Confidence Index (CCI), and Purchasing Managers' Index (PMI) \citep{FJ-Pehrsson-2009-IntMarketRev,FJ-Huang-Fang-Miller-2014-JEmpirFinanc}, inflationary pressures including Consumer Price Index (CPI) and Producer Price Index (PPI) \citep{FJ-Wu-Min-Wen-2023-EurJFinanc,FJ-Wei-Wang-Zhou-Shang-Ren-2024-FinancResLett}, and liquidity and credit conditions including broad measure of money supply (M2) and Aggregate Financing to the Real Economy (AFRE) \citep{FJ-Obstfeld-Shambaugh-Taylor-2010-AmEconJ-Macroecon,FJ-He-Wei-2023-AnnuRevEcon}.

\section{Data description}
\label{Sec_Data}

In this paper, China's financial market is divided into five sub-markets: monetary, stock, bond, foreign exchange, and bulk commodity. The monetary market is represented by the overnight (D), weekly (W), monthly (M), three-month (3M), six-month (6M), nine-month (9M), and one-year (Y) Shanghai Interbank Offered Rates (SHIBORs). The stock market is represented by the CSI 300 Index (CSI), the bond market by the China Bond Net Index (CBNI), and the foreign exchange market by the exchange rates of the U.S. dollar (US), euro (EU), Hong Kong dollar (HK), 100 Japanese yen (JP), British pound (BP), Australian dollar (AU), Canadian dollar (CD), New Zealand dollar (NZ), and Singapore dollar (SG) against the RMB. The bulk commodity market is represented by indices for Precious Metals (PM), Non-Ferrous Metals (NF), Energy (Ene), Chemicals (Che), Fats \& Oils (F\&O), Soft Commodities (SC), and Grains (Gra).

The selection of the external shock variable is essential. Existing studies have shown that the risk captured by the Chicago Board Options Exchange Volatility Index (VIX) is the primary external factor representing global financial risk \citep{FJ-Chen-Sun-2022-NAmEconFinanc} and induces volatility in China's financial market \citep{FJ-Chen-Sun-Zhang-2025-JFinancStab}. Accordingly, the VIX serves as the proxy for the external shock. All data are obtained from the Wind database: \url{www.wind.com.cn}.

To reduce noise and avoid overfitting, we uniformly employ monthly-frequency data (or monthly averages). Given data availability, the sample period spans from January 2008 to December 2023. All non-stationary series are transformed using logarithmic differencing. As shown in Table~\ref{Table_Description}, the standard deviations of the overnight, weekly, and monthly SHIBOR and the VIX exceed 0.1, suggesting more pronounced volatility. Moreover, the ADF test confirms that all series are stationary after transformation, satisfying a key prerequisite for the TVP-VAR estimation.

\begin{table}[htb!]
\renewcommand\arraystretch{0.8}
\renewcommand{\tabcolsep}{4mm}
\caption{Descriptive statistics of all series.}
\label{Table_Description}
{\centering
    \begin{tabular}{llrrrrrrrrr}
    \toprule
        ~ & Indicators & Mean & Sd & Skewness & Kurtosis & ADF \\ 
        \midrule
        \multirow{7}*{Monetary} & D & 0.0178 & 0.2003 & 1.5282 & 7.2040 & $-$6.6397$^{***}$  \\ 
        ~ & W & 0.0084 & 0.1583 & 2.1553 & 12.0019 & $-$5.4983$^{***}$  \\ 
        ~ & M & 0.0064 & 0.1487 & 2.0395 & 10.0889 & $-$5.1907$^{***}$  \\ 
        ~ & 3M & $-$0.0001 & 0.0901 & 0.0077 & 4.6822 & $-$5.1568$^{***}$  \\ 
        ~ & 6M & $-$0.0015 & 0.0719 & $-$0.7313 & 6.8441 & $-$5.5111$^{***}$  \\ 
        ~ & 9M & $-$0.0019 & 0.0656 & $-$0.8354 & 7.9341 & $-$5.4189$^{***}$  \\ 
        ~ & Y & $-$0.0022 & 0.0607 & $-$0.8648 & 8.5151 & $-$5.4818$^{***}$  \\ 
        \midrule
        Stock & CSI & $-$0.0001 & 0.0617 & 0.2438 & 2.1630 & $-$5.2294$^{***}$  \\
        \midrule
        Bond & CBNI & 0.0005 & 0.0059 & 0.2635 & 3.2576 & $-$5.6503$^{***}$  \\ 
        \midrule
        \multirow{9}*{Exchange} & US & 0.0000 & 0.0092 & 1.1075 & 4.4588 & $-$6.0040$^{***}$  \\ 
        ~ & EU & $-$0.0014 & 0.0204 & $-$0.2627 & 1.4808 & $-$5.2781$^{***}$  \\ 
        ~ & HK & 0.0000 & 0.0091 & 1.0311 & 4.4110 & $-$6.1255$^{***}$  \\ 
        ~ & JP & $-0.0018$ & 0.0222 & 0.0236 & 0.7996 & $-$6.1190$^{***}$  \\ 
        ~ & BP & $-$0.0020 & 0.0204 & $-$0.6208 & 2.1571 & $-$5.8878$^{***}$  \\ 
        ~ & AU & $-$0.0010 & 0.0272 & $-$1.2641 & 7.3526 & $-$5.2341$^{***}$  \\ 
        ~ & CD & $-$0.0014 & 0.0184 & $-$0.6991 & 5.1681 & $-$5.6060$^{***}$  \\ 
        ~ & NZ & $-$0.0009 & 0.0260 & $-$0.1385 & 1.5724 & $-$4.7893$^{***}$  \\ 
        ~ & SG & 0.0003 & 0.0111 & $-$0.0187 & 0.4640 & $-$5.0229$^{***}$  \\ 
        \midrule
        \multirow{7}*{Bulk} & PM & 0.0042 & 0.0419 & 0.2296 & 1.4285 & $-$4.9300$^{***}$  \\ 
        ~ & NF & 0.0022 & 0.0530 & $-$0.7901 & 4.8014 & $-$6.1621$^{***}$  \\ 
        ~ & Ene & 0.0021 & 0.0671 & $-$0.6218 & 3.1629 & $-$6.4814$^{***}$  \\ 
        ~ & Che & 0.0004 & 0.0647 & $-$0.4496 & 2.6755 & $-$6.9784$^{***}$  \\ 
        ~ & F\&O & 0.0001 & 0.0423 & $-$0.7338 & 3.3298 & $-$6.3096$^{***}$  \\ 
        ~ & SC & 0.0003 & 0.0410 & 0.9583 & 4.5301 & $-$5.9802$^{***}$  \\ 
        ~ & Gra & 0.0005 & 0.0255 & $-$0.3735 & 1.7268 & $-$4.9421$^{***}$  \\
        \midrule
        Shock & VIX & 0.0187 & 0.2538 & 3.7297 & 21.3701 & $-$6.2775$^{***}$  \\ 
        \bottomrule
    \end{tabular}
}
\begin{flushleft}
\footnotesize
Notes: ADF refers to the Augmented Dickey-Fuller test statistic. The asterisks $^{***}$ indicate significance at the 1\% level.
\end{flushleft}
\end{table}

\section{Empirical analysis}
\label{Sec_Empirical}

\subsection{Obtaining impulse response functions based on the TVP-VAR model}

For the TVP-VAR model, we employ Markov Chain Monte Carlo (MCMC) sampling to generate 10,000 draws, discarding the first 1,000 iterations as burn-in \citep{FJ-Tang-Liu-Zhou-2022-JIntFinancMarkInstMoney}. After obtaining the impulse response functions for all series within each sub-market, we apply principal component analysis (PCA) to construct a representative indicator for each sub-market. These indicators are then aggregated to derive a comprehensive measure for the total financial market. Table~\ref{Table_Description_TVPVAR} presents the model-fitting results for all markets. 

\begin{figure}[htb!]
\centering
\includegraphics[width=1\textwidth]{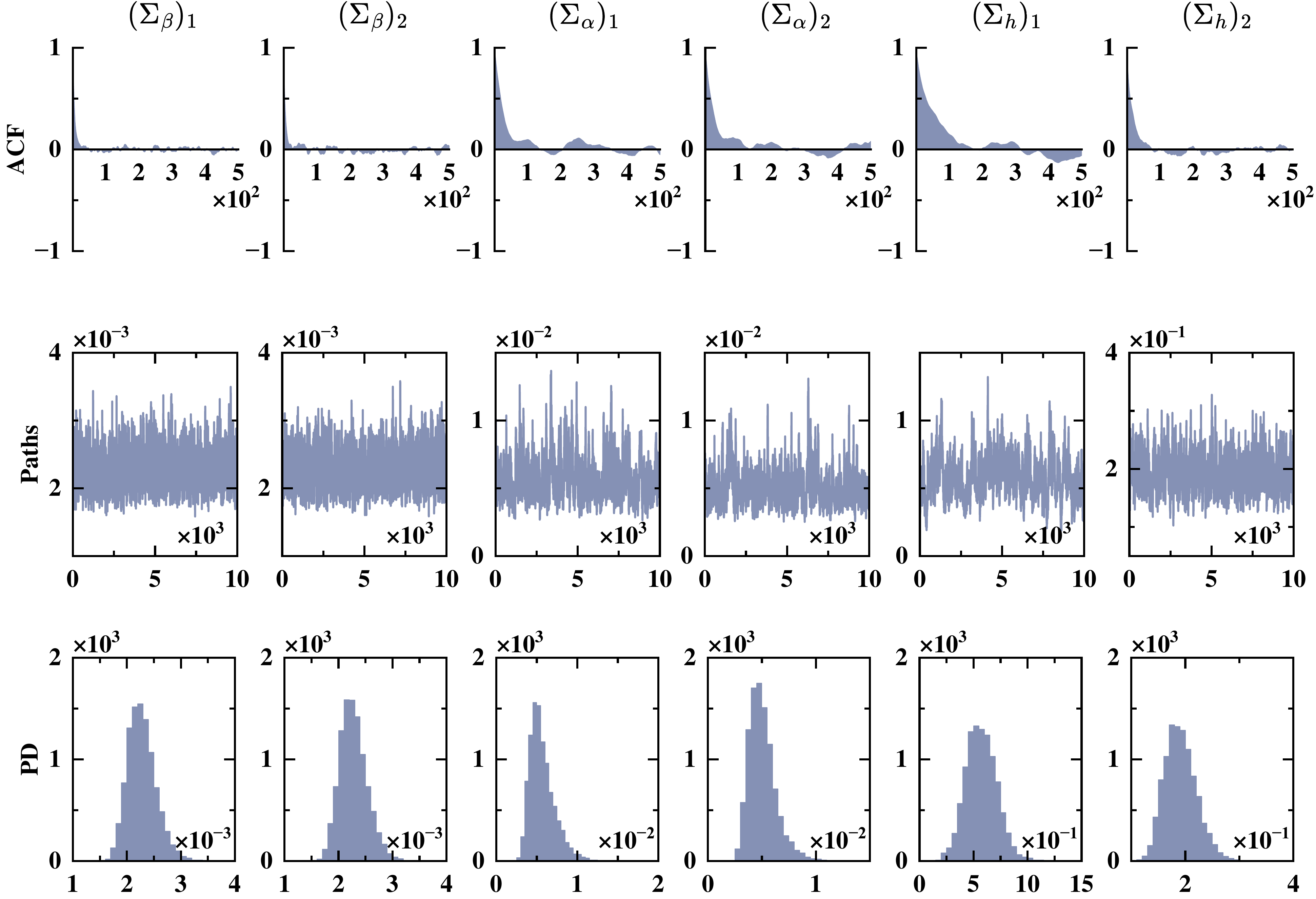}
\caption{Sample autocorrelation functions (top), sample paths (middle), and posterior distributions (bottom) for MCMC sampling. \normalfont{In this figure, ACF, Paths, and PD refer to autocorrelation functions, sample paths, and posterior distributions of six parameters, respectively.}}
\label{Fig_MCMC}
\end{figure}

Based on the CD statistic, the null hypothesis of convergence to the posterior distribution cannot be rejected at the 5\% confidence level, and the posterior means of all parameters lie within their respective 95\% credible intervals. Moreover, the inefficiency factors are all below 100, indicating that both the sampling and model estimation are reliable. Fig.~\ref{Fig_MCMC} illustrates the MCMC diagnostics for the total market, including the sample autocorrelation functions, sample paths, and posterior distributions. The results reveal that the sample paths are stable and the autocorrelations decline smoothly, confirming that the MCMC procedure effectively produces samples with low serial dependence.

\begin{table}[htb!]
\renewcommand\arraystretch{0.8}
\renewcommand{\tabcolsep}{3.7mm}
\caption{Descriptive statistics of parameters for the TVP-VAR model.}
\label{Table_Description_TVPVAR}
{\centering
    \begin{tabular}{llrrrrrr}
    \toprule
        Market & Parameters & Mean & Sd & 95\% Intervals & CD & Ineff \\ 
        \midrule
        \multirow{6}*{Total} & $(\Sigma_\beta)_1$ & 0.0023 & 0.0003 & [0.0018,0.0029] & 0.0367 & 12.3138 \\ 
        ~ & $(\Sigma_\beta)_2$ & 0.0023 & 0.0003 & [0.0018,0.0028] & 0.7829 & 11.6978 \\ 
        ~ & $(\Sigma_\alpha)_1$ & 0.0057 & 0.0016 & [0.0035,0.0094] & 0.6740 & 55.4536 \\ 
        ~ & $(\Sigma_\alpha)_2$ & 0.0051 & 0.0013 & [0.0032,0.0082] & 0.3305 & 56.9198 \\ 
        ~ & $(\Sigma_h)_1$ & 0.5723 & 0.1443 & [0.3086,0.8676] & 0.2555 & 89.1437 \\ 
        ~ & $(\Sigma_h)_2$ & 0.1911 & 0.0296 & [0.1410,0.2554] & 0.0048 & 27.6405 \\ 
        \midrule
        \multirow{6}*{Monetary} & $(\Sigma_\beta)_1$ & 0.0022 & 0.0003 & [0.0018,0.0028] & 0.9952 & 11.8438 \\ 
        ~ & $(\Sigma_\beta)_2$ & 0.0023 & 0.0003 & [0.0018,0.0029] & 0.6893 & 9.1121 \\ 
        ~ & $(\Sigma_\alpha)_1$ & 0.0322 & 0.0106 & [0.0166,0.0567] & 0.0417 & 59.7705 \\ 
        ~ & $(\Sigma_\alpha)_2$ & 0.0060 & 0.0022 & [0.0034,0.0108] & 0.9547 & 84.0997 \\ 
        ~ & $(\Sigma_h)_1$ & 0.6867 & 0.1242 & [0.4633,0.9464] & 0.4162 & 36.6596 \\ 
        ~ & $(\Sigma_h)_2$ & 0.3249 & 0.0409 & [0.2531,0.4141] & 0.0300 & 24.7197 \\ 
        \midrule
        \multirow{6}*{Stock} & $(\Sigma_\beta)_1$ & 0.0022 & 0.0004 & [0.0017,0.0030] & 0.2355 & 16.0314 \\ 
        ~ & $(\Sigma_\beta)_2$ & 0.0023 & 0.0004 & [0.0017,0.0032] & 0.0749 & 18.4634 \\ 
        ~ & $(\Sigma_\alpha)_1$ & 0.0057 & 0.0017 & [0.0035,0.0100] & 0.8372 & 52.7288 \\ 
        ~ & $(\Sigma_\alpha)_2$ & 0.0055 & 0.0015 & [0.0033,0.0093] & 0.0061 & 62.5183 \\ 
        ~ & $(\Sigma_h)_1$ & 0.0062 & 0.0021 & [0.0035,0.0116] & 0.0018 & 95.8387 \\ 
        ~ & $(\Sigma_h)_2$ & 0.5233 & 0.1211 & [0.3057,0.7794] & 0.7402 & 59.7017 \\ 
        \midrule
        \multirow{6}*{Bond} & $(\Sigma_\beta)_1$ & 0.0020 & 0.0003 & [0.0016,0.0026] & 0.8593 & 13.5840 \\ 
        ~ & $(\Sigma_\beta)_2$ & 0.0023 & 0.0004 & [0.0017,0.0032] & 0.3238 & 18.2550 \\ 
        ~ & $(\Sigma_\alpha)_1$ & 0.0055 & 0.0018 & [0.0035,0.0100] & 0.3477 & 57.4730 \\ 
        ~ & $(\Sigma_\alpha)_2$ & 0.0055 & 0.0016 & [0.0034,0.0095] & 0.0049 & 54.9616 \\ 
        ~ & $(\Sigma_h)_1$ & 0.3899 & 0.0432 & [0.3140,0.4810] & 0.1926 & 35.9936 \\ 
        ~ & $(\Sigma_h)_2$ & 0.5501 & 0.1133 & [0.3304,0.7845] & 0.1736 & 45.4013 \\ 
        \midrule
        \multirow{6}*{Exchange} & $(\Sigma_\beta)_1$ & 0.0023 & 0.0003 & [0.0018,0.0028] & 0.9512 & 8.6598 \\ 
        ~ & $(\Sigma_\beta)_2$ & 0.0023 & 0.0003 & [0.0018,0.0028] & 0.6862 & 8.7490 \\ 
        ~ & $(\Sigma_\alpha)_1$ & 0.0055 & 0.0016 & [0.0034,0.0095] & 0.1155 & 44.5438 \\ 
        ~ & $(\Sigma_\alpha)_2$ & 0.0053 & 0.0015 & [0.0033,0.0090] & 0.6252 & 56.4034 \\ 
        ~ & $(\Sigma_h)_1$ & 0.5378 & 0.0543 & [0.4464,0.6570] & 0.8152 & 33.0456 \\ 
        ~ & $(\Sigma_h)_2$ & 0.2609 & 0.0325 & [0.2045,0.3333] & 0.1404 & 18.0483 \\ 
        \midrule
        \multirow{6}*{Bulk} & $(\Sigma_\beta)_1$ & 0.0023 & 0.0002 & [0.0018,0.0028] & 0.6945 & 7.5160 \\ 
        ~ & $(\Sigma_\beta)_2$ & 0.0023 & 0.0002 & [0.0018,0.0028] & 0.2311 & 7.6802 \\ 
        ~ & $(\Sigma_\alpha)_1$ & 0.0054 & 0.0016 & [0.0034,0.0092] & 0.5347 & 78.2338 \\ 
        ~ & $(\Sigma_\alpha)_2$ & 0.0057 & 0.0017 & [0.0035,0.0099] & 0.0001 & 65.4734 \\ 
        ~ & $(\Sigma_h)_1$ & 0.0057 & 0.0016 & [0.0034,0.0096] & 0.0523 & 54.8941 \\ 
        ~ & $(\Sigma_h)_2$ & 0.2035 & 0.0272 & [0.1578,0.2643] & 0.9039 & 39.1602 \\ 
        \bottomrule
    \end{tabular}
}
\begin{flushleft}
\footnotesize
Notes: CD refers to the Convergence Diagnostics used to test the convergence of Markov Chains. Ineff refers to the ineffective factor. See \cite{FJ-Geweke-1992-BayesStat} for more information.
\end{flushleft}
\end{table}

\begin{figure}[htb!]
\centering
\includegraphics[width=1\textwidth]{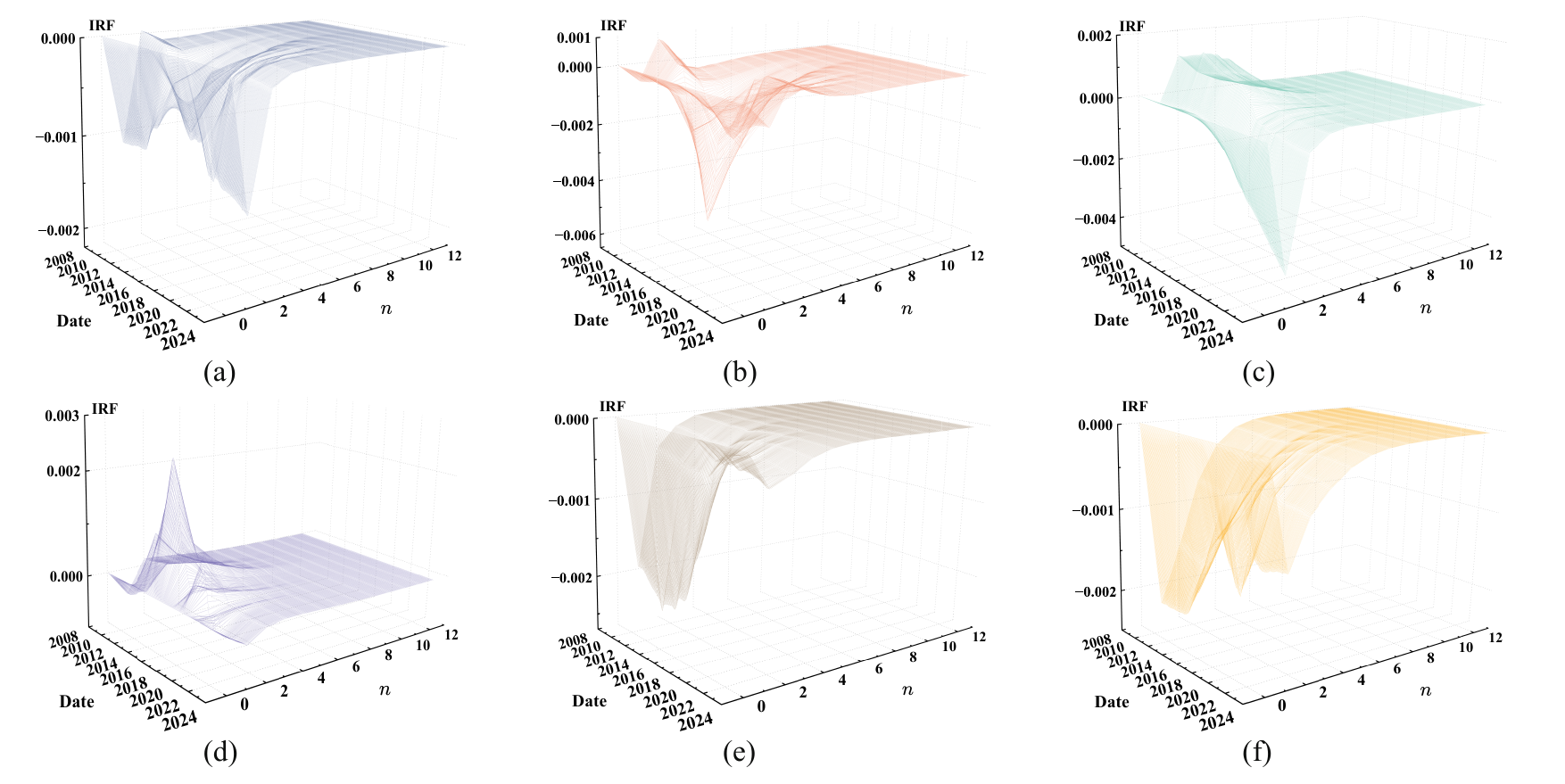}
\caption{The evolution of impulse response functions of markets to the VIX. \normalfont{In this figure, $n$ refers to the number of periods of the response function. (a) Total market; (b) Monetary market; (c) Stock market; (d) Bond market; (e) Foreign exchange market; (f) Bulk commodity market.}}
\label{Fig_IRF3D} 
\end{figure}

Fig.~\ref{Fig_IRF3D} presents the impulse responses of the six market indicators to VIX shocks. The stock and bulk commodity markets exhibit stronger and more persistent responses, whereas the monetary, bond, and foreign exchange markets appear relatively resilient, consistent with the findings of \cite{FJ-Chen-Sun-2022-NAmEconFinanc}. The directions of these responses also differ across markets. For instance, between 2008 and 2014, VIX shocks had positive short-run effects on the bond market (within approximately three months). These findings suggest that risky assets are more prone to panic-induced selloffs, whereas safe-haven assets serve a stabilizing function \citep{FJ-Cheema-Faff-Szulczyk-2022-IntRevFinancAnal}. During these periods, capital may flow out of risky assets and into highly rated bonds, such as Treasuries, thereby exerting a positive influence on the bond market. Moreover, the convergence speeds of the IRFs vary across markets. For example, in 2015, the monetary market's IRFs converged to zero within twelve months, while the bond market's responses dissipated within five months, indicating heterogeneity in both the speed and persistence of market adjustments to shocks. This observed heterogeneity in response dynamics motivates our separate measurement of adaptability and recoverability, providing a richer characterization of market resilience.

\subsection{Quantifying China's financial market resilience}

\begin{figure}[htb!] 
  \centering  
  \includegraphics[width=1\textwidth]{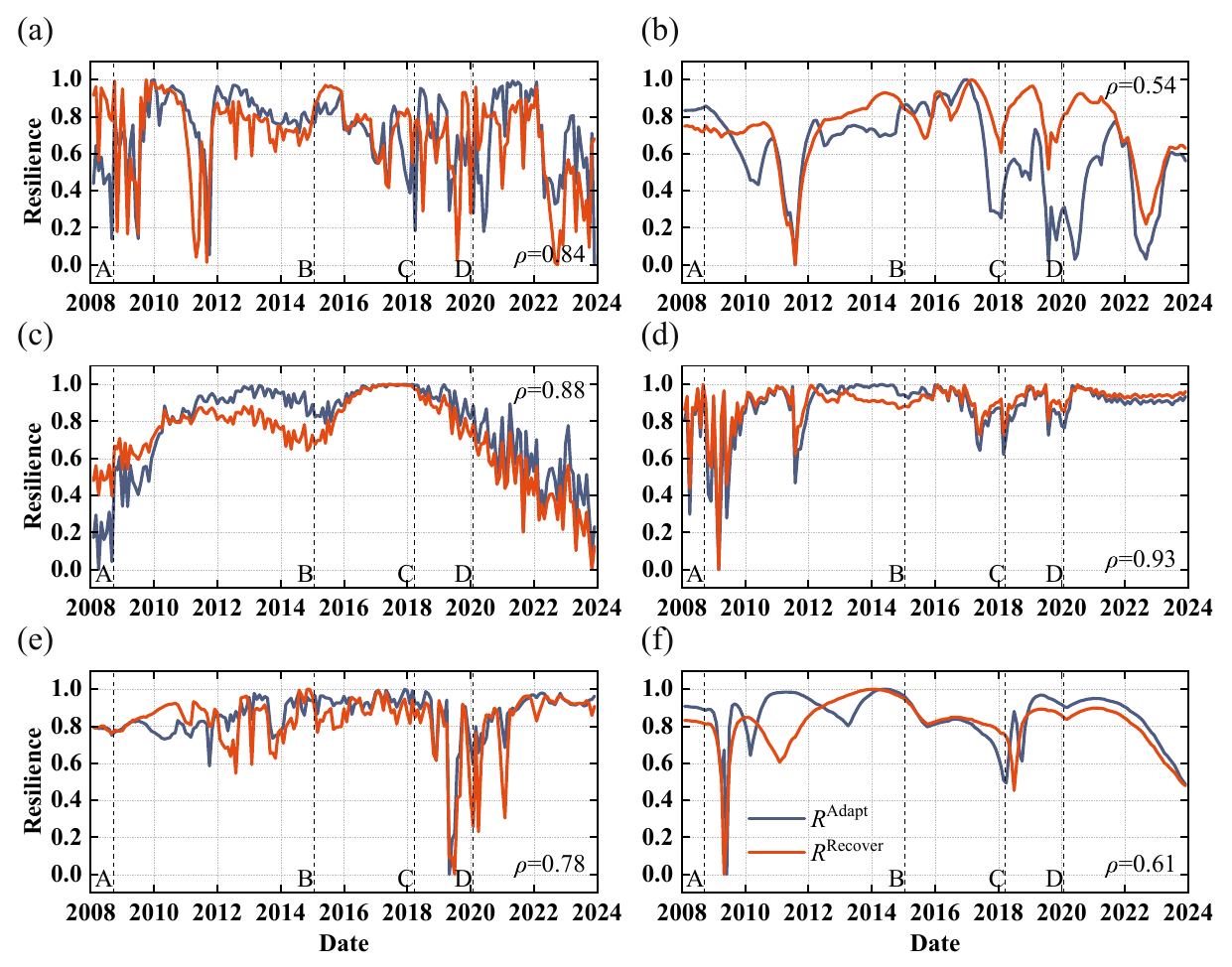} 
  \caption{The evolution of resilience for China's financial markets. \normalfont{In this figure, the black dotted lines indicate the timing of major events: A, September 15, 2008, the bankruptcy of Lehman Brothers; B, January 19, 2015, a 7.7\% plunge in China's Shanghai Stock Exchange Index; C, March 22, 2018, President Trump's memorandum imposing tariffs on Chinese imports; and D, January 23, 2020, Wuhan's lockdown announcement in response to the COVID-19 pandemic. $\rho$ refers to the correlation coefficient between $R^\mathrm{Adapt}$ and $R^\mathrm{Recover}$. (a) Total market; (b) Monetary market; (c) Stock market; (d) Bond market; (e) Foreign exchange market; (f) Bulk commodity market.}}
  \label{Fig_FinResilience_AdaRec} 
\end{figure}

Based on the time-varying impulse response estimates from the TVP-VAR model, we construct financial market resilience indices for China's total market and its five sub-markets. These indices are derived according to Eqs.~(\ref{eq_adaptability}) and (\ref{eq_recoverability}), which measure adaptability and recoverability, respectively, as illustrated in Fig.~\ref{Fig_FinResilience_AdaRec}. To ensure temporal comparability, all indices are normalized, and the shock duration $N$ is fixed at 12 months \citep{FJ-Tang-Liu-Zhou-2022-JIntFinancMarkInstMoney}. Overall, the results indicate that $R^\mathrm{Adapt}$ for each market is strongly and positively correlated with $R^\mathrm{Recover}$, particularly in the stock and bond markets ($\rho = 0.88$ and $\rho = 0.93$), suggesting that stronger risk absorption capacity is associated with greater recovery ability \citep{FJ-Tang-Liu-Zhou-2022-JIntFinancMarkInstMoney}. In addition, the resilience of China's financial markets exhibits distinct event-driven patterns, as shocks from the four major events (A-D) consistently trigger pronounced short-term declines in resilience levels.

Since the end of 2008, the resilience of China's financial markets deteriorated, particularly in the total and bond markets. This trend is mainly attributed to the outbreak of the global financial crisis and China's ``four trillion yuan'' economic stimulus plan. On the one hand, the crisis triggered global market panic and a sustained rise in the VIX, which undermined the stability of China's financial markets \citep{FJ-Chen-Sun-2022-NAmEconFinanc}. On the other hand, although the stimulus plan\textemdash characterized by an active fiscal policy and a moderately loose monetary policy\textemdash played a crucial role in stimulating social investment and stabilizing the domestic economy, it also caused a sharp increase in local government debt across provinces and regions \citep{FJ-Zou-2024-EconModel}. Consequently, financial system instability intensified, as reflected in the rapid expansion of shadow banking in subsequent years \citep{FJ-Chen-He-Liu-2020-JFinancEcon}, thereby weakening the resilience of China's bond market and, in turn, the overall financial market.

Subsequently, the resilience declined sharply between 2011 and 2013, during which the European sovereign debt crisis intensified, and some international capital flowed back from emerging markets as a precautionary measure. This affected China's cross-border capital movements, exacerbated the dollar shortage, and increased pressure on domestic banking liquidity, as clearly shown in Fig.~\ref{Fig_FinResilience_AdaRec}b. Shortly afterward, in April 2014, China's financial supervisory authority introduced the new ``Nine Principles'', which prompted a continuous inflow of social capital into the stock market, resulting in abundant liquidity. Combined with positive official statements and accommodative policy stances, these factors fueled two rapid surges in China's stock prices \citep{FJ-Liu-Gu-Xing-2016-IntRevEconFinanc}. However, as leverage and market risks accumulated, regulators' intensified investigations into margin financing triggered the onset of a major bear market \citep{FJ-Li-Zheng-Liu-2022-RegulGov}. Since June 2015, China's stock market experienced three large-scale crashes, with the stock index plunging by 49\% within six months and market capitalization shrinking by approximately 36 trillion yuan.

Despite that, it was not evident that the resilience continued to decline or fluctuate, for two main reasons. First, the Chinese government implemented coordinated rescue measures in the secondary market to stabilize investor confidence, and as a result, the crash did not cause significant damage to the real economy due to timely interventions \citep{FJ-Chen-Gong-2019-IntRevEconFinanc}. Second, as an endogenous, localized shock, the source of risk in this crash was concentrated within the Chinese stock market rather than driven by global panic represented by the VIX. Consequently, since the resilience indicator in this paper relies on external shocks, it under-identifies structural shocks originating within the market.

Afterward, the resilience gradually recovered until the onset of trade friction between China and the United States in March 2018. Trade serves as a primary determinant of the exchange rate, which in turn strongly influences interest rates and commodity prices \citep{FJ-Linnemann-Schabert-2015-JIntEcon,FJ-Gagnon-Sarsenbayev-2023-JIntMoneyFinan}. Since March 2018, the United States repeatedly restricted China's trade activities by imposing tariffs, limiting imports, and tightening financing conditions, exerting substantial pressure on China's financial markets. In the monetary market, changes in external demand and inflation expectations complicated monetary policy adjustments. In the foreign exchange market, the RMB faced depreciation pressure due to disruptions in international trade. In the bulk commodity market, prices of soybeans, corn, steel, and other commodities fluctuated sharply, squeezing corporate profit margins and threatening financial stability. Consequently, the two indicators for the foreign exchange market reached historical lows, which in turn dampened the resilience of other sub-markets. Since the end of 2019, the COVID-19 pandemic accelerated the global economic downturn, placing unprecedented pressure on China's financial market resilience. For example, in the monetary, stock, and foreign exchange markets, both indicators declined sharply, likely due to liquidity constraints in the real economy, downward adjustments in market expectations, and pressures from capital outflows.

\begin{figure}[htb!] 
  \centering  
  \includegraphics[width=1\textwidth]{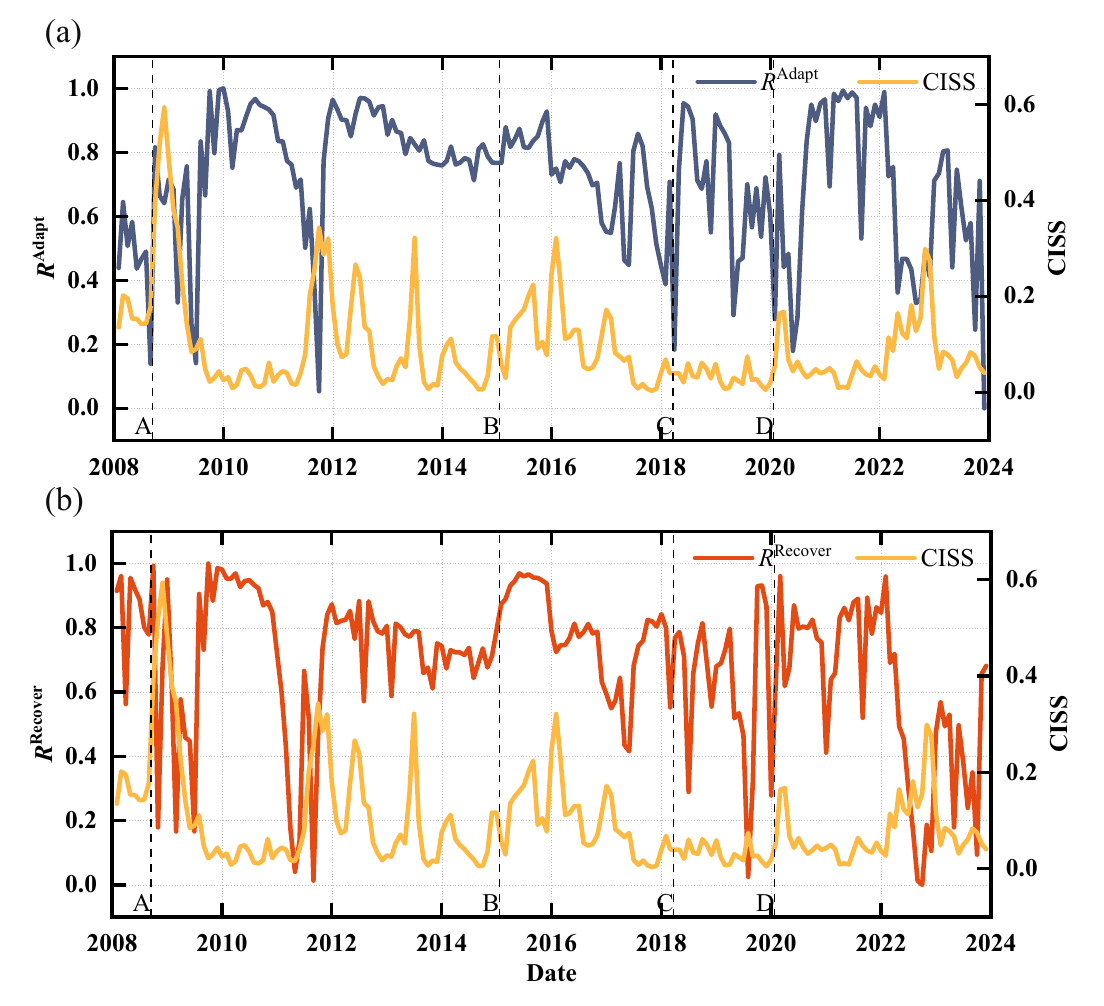} 
  \caption{The evolution of resilience for China's financial total market and its Composite Indicator of Systemic Stress (CISS). \normalfont{In this figure, the black dotted lines indicate the timing of major events, as shown in Fig.~\ref{Fig_FinResilience_AdaRec}. (a) $R^\mathrm{Adapt}$ and CISS; (b) $R^\mathrm{Recover}$ and CISS.}}
  \label{Fig_CISS_Resilience_AdaRec} 
\end{figure}

In Fig.~\ref{Fig_CISS_Resilience_AdaRec}, we also compare the evolution of China's total financial market resilience with its Composite Indicator of Systemic Stress (CISS)\footnote{\url{https://data.ecb.europa.eu/data/datasets}}, proposed by the European Central Bank \citep{FJ-Hollo-Kremer-LoDuca-2012-ECB}. The CISS comprises 16 individual financial stress sub-indices and measures systemic risk in the financial sector, with higher values indicating worse financial conditions. Notably, fluctuations in $R^\mathrm{Adapt}$ and $R^\mathrm{Recover}$ were negatively correlated with those of the CISS, particularly during the 2008 global financial crisis and the COVID-19 pandemic from 2020 to 2022, suggesting that the two indicators in this study accurately captured China's financial market resilience during major crises. Moreover, during the 2015 stock market crash, China's CISS appeared more sensitive than the indicators proposed in this paper, further highlighting that the latter primarily reflect resilience to external shocks.

\subsection{Connectedness of resilience among China's financial sub-markets}

The volatility of different asset prices strongly affects each other, and understanding such volatility spillovers is crucial for developing trading and hedging strategies, as well as for formulating regulatory policies \citep{FJ-Wang-Xie-Jiang-Stanley-2016-FinancResLett}. Analogously, investigating volatility spillovers in the resilience of China's sub-markets provides a novel perspective for risk management. Accordingly, we employ the DY connectedness approach to measure the static connectedness indices of financial market resilience among China's sub-markets, as presented in Table~\ref{Table_StaticConnectedness}.

\begin{table}[htb!]
\renewcommand\arraystretch{1.1}
\renewcommand{\tabcolsep}{5mm}
\belowrulesep=0pt
\aboverulesep=0pt
\caption{Static connectedness indices of resilience among China's financial sub-markets.}
\label{Table_StaticConnectedness}
\centering
    \begin{tabular}{lrrrrrr}
    \toprule
          & Monetary & Stock & Bond & Exchange & Bulk & $From$ \\ 
        \midrule       
        \multicolumn{6}{l}{\it{Panel A: Adaptability}} \\
        Monetary & 86.59 &  2.12 &  2.97 &  2.19 &  6.13 & 13.41  \\ 
        Stock &  3.46 & 82.15 &  5.16 &  5.77 &  3.46 & 17.85  \\ 
        Bond &  2.99 &  2.80 & 78.59 &  2.85 & 12.77 & 21.41  \\ 
        Exchange &  1.27 &  2.61 &  1.05 & 93.29 &  1.78 &  6.71  \\ 
        Bulk &  3.44 &  1.25 &  8.96 & 13.73 & 72.62 & 27.38  \\ 
        $To$ & 11.17 &  8.78 & 18.14 & 24.53 & 24.14 & \multirow{2}{*}{\makecell{$\boldsymbol{TCI=}$\\\bf 86.76}}  \\ 
        $NET$ & $-$2.24 & $-$9.07 & $-$3.27 & 17.83 & $-$3.24 & ~ \\
        \midrule
        \multicolumn{6}{l}{\it{Panel B: Recoverability}}\\
        Monetary & 84.58 &  4.23 &  2.35 &  5.81 &  3.04 & 15.42  \\ 
        Stock &  1.82 & 86.16 &  2.49 &  6.77 &  2.76 & 13.84  \\ 
        Bond &  3.91 &  6.54 & 77.42 &  3.40 &  8.73 & 22.58  \\ 
        Exchange &  1.67 &  2.29 &  0.39 & 91.68 &  3.97 &  8.32  \\ 
        Bulk &  4.01 &  1.27 &  2.92 &  2.02 & 89.79 & 10.21  \\ 
        $To$ & 11.41 & 14.33 &  8.14 & 17.99 & 18.50 & \multirow{2}{*}{\makecell{$\boldsymbol{TCI=}$\\\bf 70.38}}  \\ 
        $NET$ & $-$4.01 & 0.49 & $-$14.44 & 9.67 & 8.29 & ~ \\
    \bottomrule
    \end{tabular}
\end{table}

The $TCI$ values of adaptability and recoverability among the five markets are 86.76\% and 70.38\%, respectively, indicating that the ability of each sub-market to adapt to external financial risks and recover from them is highly correlated with that of others, making the overall system prone to volatility resonance \citep{FJ-Wang-Xie-Jiang-Stanley-2016-FinancResLett,FJ-Zhao-Zhang-Liu-2020-EmergMarkFinancTrade}. Compared with the $TCI$ of recoverability, that of adaptability is higher, indicating a stronger cascading effect in the adaptability of China's financial sub-markets and relatively less consistency in their recoverability.

According to the $NET$ results, we find significant net spillovers from the adaptability and recoverability of the foreign exchange market, highlighting its dominant role as a transmitter of volatility relative to other markets. This finding aligns with \cite{FJ-Chen-Mo-Qin-Yang-2023-Pac-BasinFinancJ}: Supported by the flow-oriented model \citep{FJ-Dornbusch-Fischer-1980-AmEconRev}, changes in exchange rates may alter currency liquidity, firms' international competitiveness, and bulk commodity trade, thereby ultimately affecting market volatility \citep{FJ-Qiao-Ding-Han-Li-2024-JIntMoneyFinan}, including financial resilience.

We also find that China's stock market primarily acts as a net recipient of spillovers, with $NET$ values for adaptability and recoverability of $-9.07\%$ and $0.49\%$, respectively. For instance, in the bulk commodity market, rapid price fluctuations, high trading volumes, and cyclical movements generate net volatility spillovers toward the stock market, reflecting the deepening financialization of the bulk commodity sector \citep{FJ-Wen-Cao-Liu-Wang-2021-IntRevFinancAnal}. Meanwhile, the results reveal asymmetry in the $NPDC$ of China's financial sub-markets, highlighting the importance for regulators to identify the sources of volatility.

\begin{figure}[htb!]
\centering
\includegraphics[width=1\textwidth]{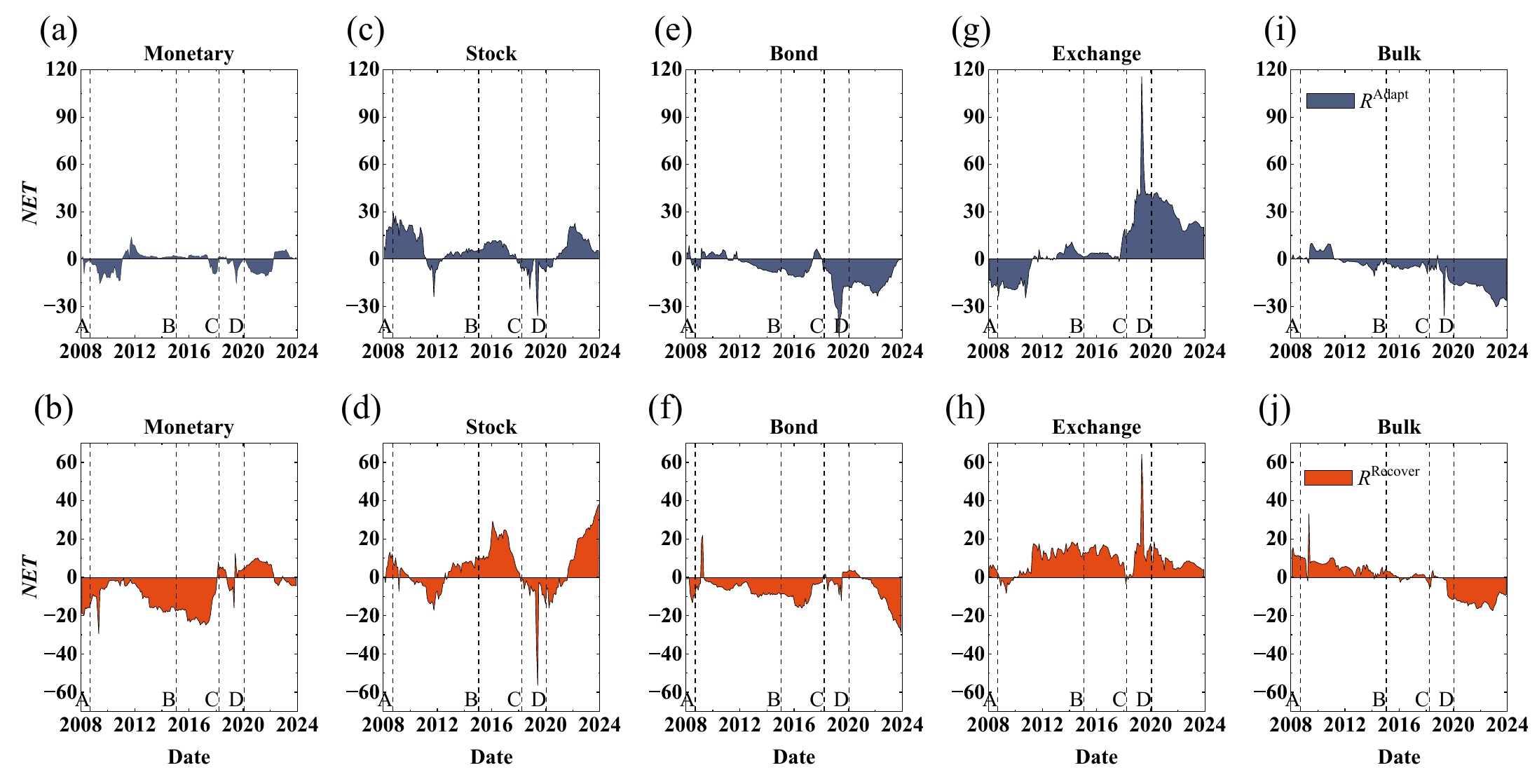}
\caption{Dynamic net connectedness of resilience among China's financial sub-markets. \normalfont{In this figure, the black dotted lines indicate the timing of major events, as shown in Fig.~\ref{Fig_FinResilience_AdaRec}.}}
\label{Fig_NET_submarkets}
\end{figure}

\begin{figure}[htb!]
\centering
\includegraphics[width=1\textwidth]{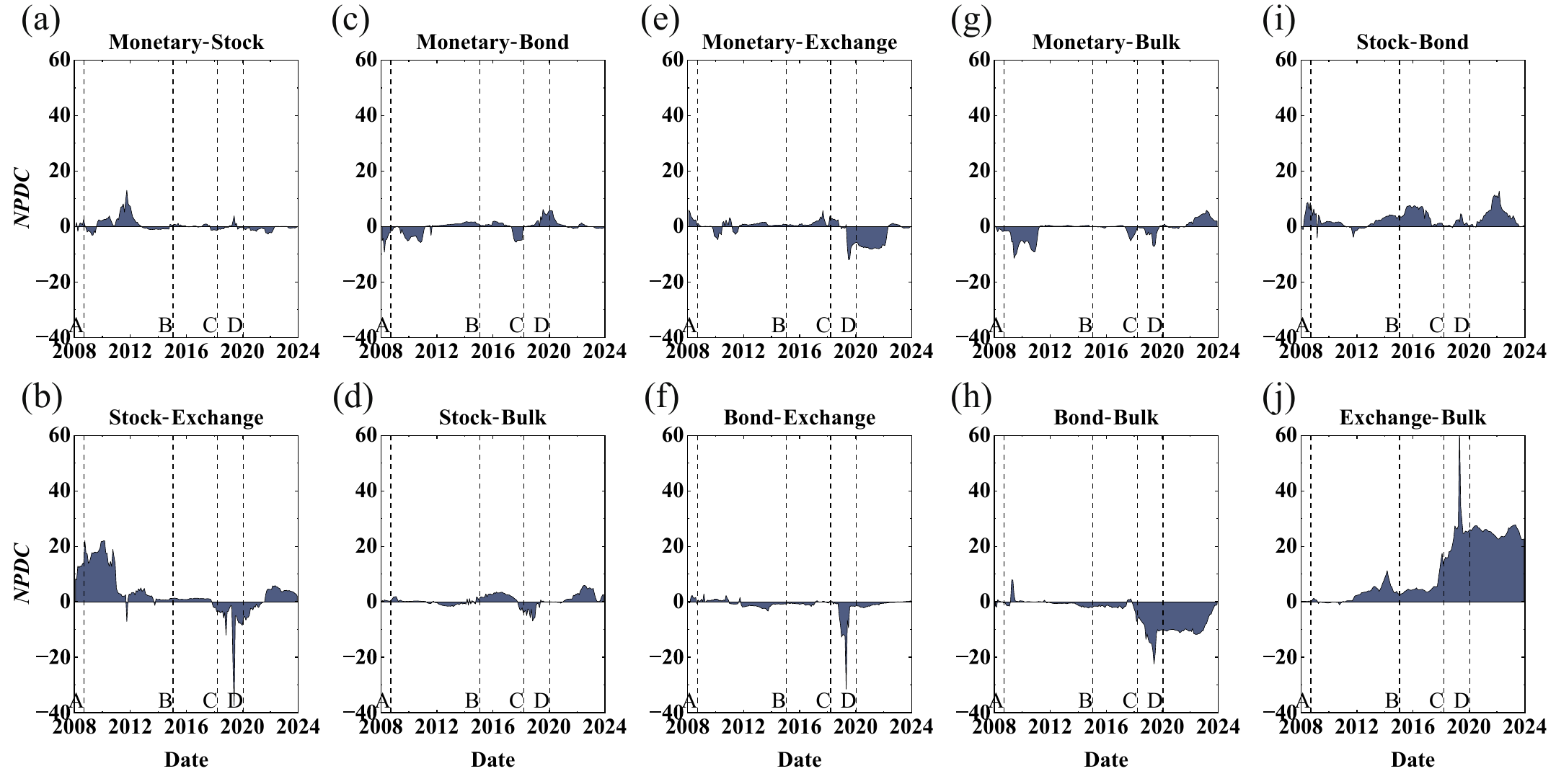}
\caption{Dynamic net pairwise directional connectedness of $R^\text{Adapt}$ among China's financial sub-markets. \normalfont{In this figure, the black dotted lines indicate the timing of major events, as shown in Fig.~\ref{Fig_FinResilience_AdaRec}.}}
\label{Fig_NPDC_submarkets_Ada}
\end{figure}

\begin{figure}[htb!]
\centering
\includegraphics[width=1\textwidth]{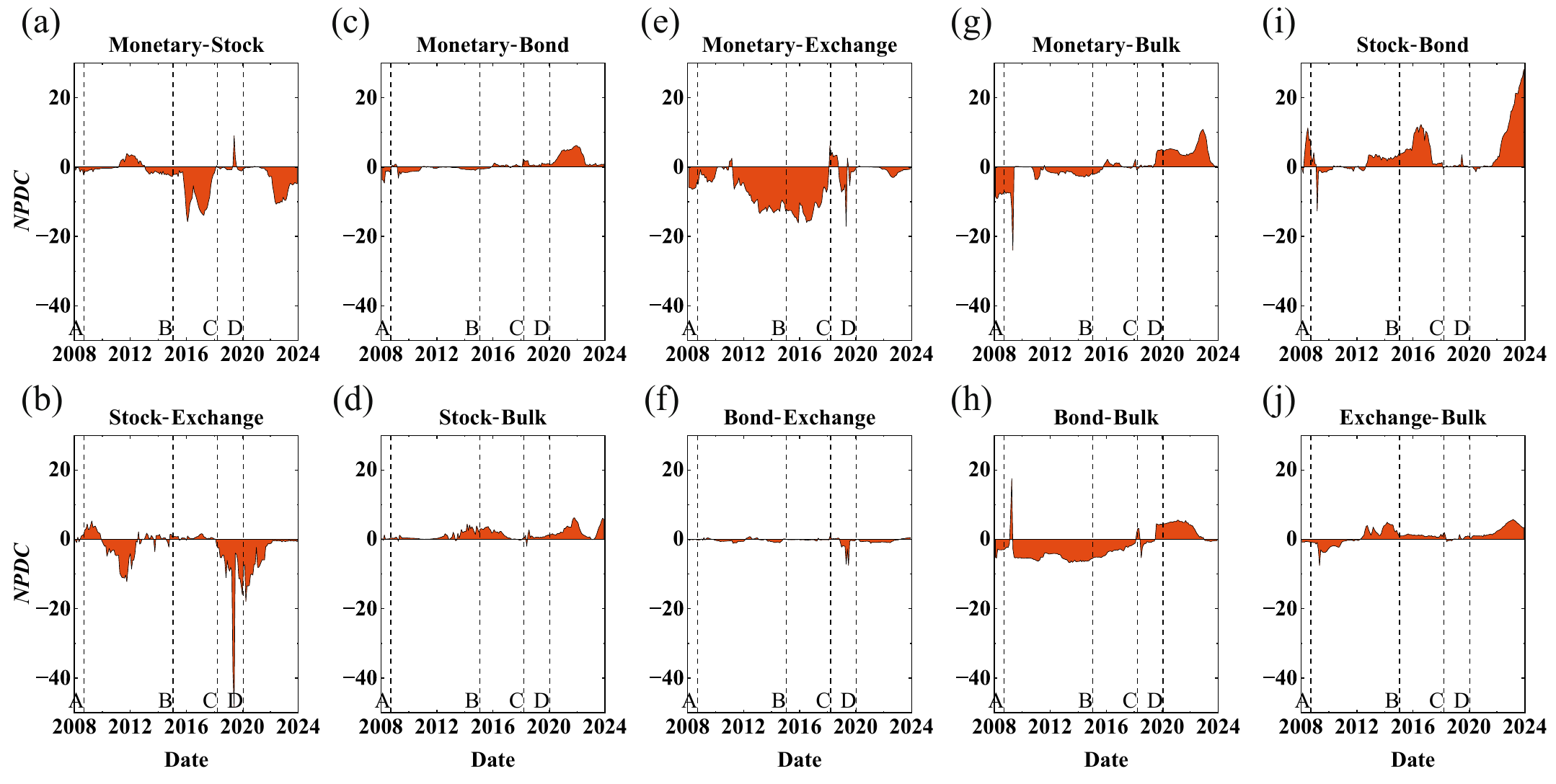}
\caption{Dynamic net pairwise directional connectedness of $R^\text{Recover}$ among China's financial sub-markets. \normalfont{In this figure, the black dotted lines indicate the timing of major events, as shown in Fig.~\ref{Fig_FinResilience_AdaRec}.}}
\label{Fig_NPDC_submarkets_Rec}
\end{figure}

While the static analysis in Table~\ref{Table_StaticConnectedness} provides a valuable overview of the average connectedness, it may mask significant time-varying patterns. To uncover how these spillover dynamics evolve, particularly during crisis periods, we present their dynamic $NET$ values based on the TVP-VAR-DY model, as shown in Fig.~\ref{Fig_NET_submarkets}. Similarly, their dynamic $NPDC$ values are presented in Figs.~\ref{Fig_NPDC_submarkets_Ada} and \ref{Fig_NPDC_submarkets_Rec}, respectively.

The results reveal evolving spillover roles among China's financial sub-markets, particularly during major events. During the 2015 stock market crash, the stock market's volatility spillover role became pronounced in both adaptability and recoverability dimensions (Fig.~\ref{Fig_NET_submarkets}c-d), in contrast to its role as a recipient of spillbacks during the 2018 US-Sino trade conflict. Specifically, we observe strong connectedness from the stock market to the bond market (Fig.~\ref{Fig_NPDC_submarkets_Ada}i and Fig.~\ref{Fig_NPDC_submarkets_Rec}i), highlighting a key transmission channel of systemic risk at this time \citep{FJ-Sun-Wu-Zeng-Peng-2021-FinancResLett}. Moreover, the bulk commodity market exhibited significant spillovers after 2008 and, in turn, pronounced spillbacks following the 2018 trade conflict (Fig.~\ref{Fig_NET_submarkets}i-j). During this period, all sub-markets acted as net recipients of spillovers, except for the foreign exchange market, whose $NET$ values for adaptability and recoverability reached nearly 120\% and 60\%, respectively (Fig.~\ref{Fig_NET_submarkets}g-h), underscoring its central role as the source of systemic risk \citep{FJ-He-Liang-Liu-2024-JIntMoneyFinan,FJ-Chen-Mo-Qin-Yang-2023-Pac-BasinFinancJ}.

In addition, clear heterogeneity emerges between adaptability- and recoverability-based spillovers. The monetary market absorbed significant spillovers from the recoverability of the stock and foreign exchange markets but responded weakly to their adaptability, suggesting that recoverability served as the dominant contagion channel. Outflows of risky capital from unrecovered markets triggered liquidity shocks and credit tensions, further weakening the monetary market's recoverability \citep{FJ-Lu-Bessler-Leatham-2018-JIntFinancMarkInstMoney}.

\subsection{Impact of uncertainty on resilience of China's financial market}

Having established the dynamics of financial resilience and its cross-market spillovers, we now address the central question of this paper: how do various forms of uncertainty affect this resilience? Firstly, Table~\ref{Table_Regression_Total} reports the baseline regression results of five external shocks on the resilience of China's total financial market. Regardless of whether adaptability or recoverability is considered, CGPR, CEPU, UCT, and CTPU all exerted significant negative effects on financial resilience, and these results remained robust after including control variables. By contrast, CCPU had a statistically significant negative impact only on recoverability, suggesting that its influence on adaptability, and thus on overall resilience, was limited.

Compared with other types of uncertainty, climate policy shocks are typically long-term and progressive, characterized by gradual transitions and extended adjustment periods that allow markets to adapt \citep{FJ-Berestycki-Carattini-DechezleprEtre-Kruse-2022-OECD,FJ-Liang-Goodell-Li-2024-JIntFinancMarkInstMoney}. In the short run, shifts in climate policy likely primarily affected high-carbon and energy-intensive sectors \citep{FJ-Han-Sun-Jiang-2024-FrontEnvironSci}. Furthermore, both China's CCI and PMI contributed positively to financial market resilience, underscoring the vital role of market confidence in maintaining systemic stability \citep{FJ-Chernykh-Davydov-Sihvonen-2023-JFinancStab}.

\begin{table}[htb!]
\centering
\caption{Regression results of external shocks on resilience of China's financial total market}
\label{Table_Regression_Total}
\footnotesize
\setlength{\tabcolsep}{4pt}
\renewcommand{\tabcolsep}{1mm}
\begin{tabular}{l*{10}{c}}
\toprule
& \multicolumn{5}{c}{$R^\text{Adapt}$} & \multicolumn{5}{c}{$R^\text{Recover}$} \\
\cmidrule(lr){2-6} \cmidrule(lr){7-11}
& (1) & (2) & (3) & (4) & (5) & (6) & (7) & (8) & (9) & (10) \\
\midrule
CGPR & $-0.146^{**}$ &&&&& $-0.312^{***}$ &&&& \\
     & $(-2.006)$ &&&&& $(-4.669)$ &&&& \\
CEPU & & $-0.250^{***}$ &&&& & $-0.341^{***}$ &&& \\
     & & $(-3.234)$ &&&& & $(-4.744)$ &&& \\
CCPU & & & $0.076$ &&&& & $-0.123^{*}$ && \\
     & & & $(1.008)$ &&&& & $(-1.713)$ && \\
UCT  & & & & $-0.230^{***}$ &&&& & $-0.193^{**}$ & \\
     & & & & $(-2.876)$ &&&& & $(-2.498)$ & \\
CTPU & & & & & $-0.239^{***}$ &&&& & $-0.321^{***}$ \\
     & & & & & $(-2.808)$ &&&& & $(-4.022)$ \\
IAV  & $0.025$ & $0.009$ & $0.032$ & $0.006$ & $0.024$ & $0.057$ & $0.038$ & $0.060$ & $0.047$ & $0.059$ \\
     & $(0.340)$ & $(0.124)$ & $(0.432)$ & $(0.087)$ & $(0.331)$ & $(0.858)$ & $(0.575)$ & $(0.855)$ & $(0.668)$ & $(0.872)$ \\
CCI  & $0.132^{*}$ & $0.193^{**}$ & $0.102$ & $0.204^{***}$ & $0.235^{***}$ & $0.308^{***}$ & $0.376^{***}$ & $0.283^{***}$ & $0.342^{***}$ & $0.431^{***}$ \\
     & $(1.819)$ & $(2.570)$ & $(1.391)$ & $(2.610)$ & $(2.811)$ & $(4.616)$ & $(5.372)$ & $(4.031)$ & $(4.544)$ & $(5.480)$ \\
PMI  & $0.140^{*}$ & $0.118$ & $0.168^{**}$ & $0.101$ & $0.136^{*}$ & $0.143^{**}$ & $0.130^{*}$ & $0.174^{**}$ & $0.137^{*}$ & $0.154^{**}$ \\
     & $(1.865)$ & $(1.597)$ & $(2.239)$ & $(1.318)$ & $(1.840)$ & $(2.080)$ & $(1.878)$ & $(2.422)$ & $(1.859)$ & $(2.220)$ \\
CPI  & $-0.024$ & $-0.069$ & $-0.026$ & $-0.041$ & $-0.055$ & $0.098$ & $0.035$ & $0.090$ & $0.080$ & $0.054$ \\
     & $(-0.336)$ & $(-0.955)$ & $(-0.359)$ & $(-0.577)$ & $(-0.769)$ & $(1.488)$ & $(0.525)$ & $(1.297)$ & $(1.163)$ & $(0.798)$ \\
PPI  & $0.009$ & $0.013$ & $0.015$ & $0.023$ & $0.006$ & $-0.005$ & $0.007$ & $0.017$ & $0.018$ & $-0.001$ \\
     & $(0.118)$ & $(0.177)$ & $(0.209)$ & $(0.317)$ & $(0.089)$ & $(-0.082)$ & $(0.105)$ & $(0.248)$ & $(0.260)$ & $(-0.020)$ \\
M2   & $-0.041$ & $-0.032$ & $-0.021$ & $-0.031$ & $-0.044$ & $0.015$ & $0.030$ & $-0.003$ & $0.029$ & $0.014$ \\
     & $(-0.560)$ & $(-0.447)$ & $(-0.281)$ & $(-0.419)$ & $(-0.607)$ & $(0.226)$ & $(0.449)$ & $(-0.038)$ & $(0.415)$ & $(0.207)$ \\
AFRE  & $0.051$ & $0.071$ & $0.035$ & $0.049$ & $0.082$ & $-0.031$ & $-0.006$ & $-0.016$ & $-0.036$ & $0.009$ \\
     & $(0.681)$ & $(0.968)$ & $(0.455)$ & $(0.671)$ & $(1.105)$ & $(-0.456)$ & $(-0.081)$ & $(-0.226)$ & $(-0.51)$ & $(0.132)$ \\
\midrule
$R^2$   & $0.071$ & $0.102$ & $0.056$ & $0.092$ & $0.090$ & $0.219$ & $0.221$ & $0.139$ & $0.154$ & $0.196$ \\
$R^2$-adj  & $0.029$ & $0.062$ & $0.014$ & $0.051$ & $0.049$ & $0.184$ & $0.187$ & $0.101$ & $0.117$ & $0.161$ \\
\bottomrule
\end{tabular}
\begin{flushleft}
\footnotesize
Notes: $R^2$-adj refers to the adjusted $R^2$. Values in parentheses represent t-values. The asterisks $^{*}$, $^{**}$, and $^{***}$ indicate significance at the 10\%, 5\%, and 1\% level, respectively.
\end{flushleft}
\end{table}

\begin{figure}[htb!]
\centering
\includegraphics[width=1\textwidth]{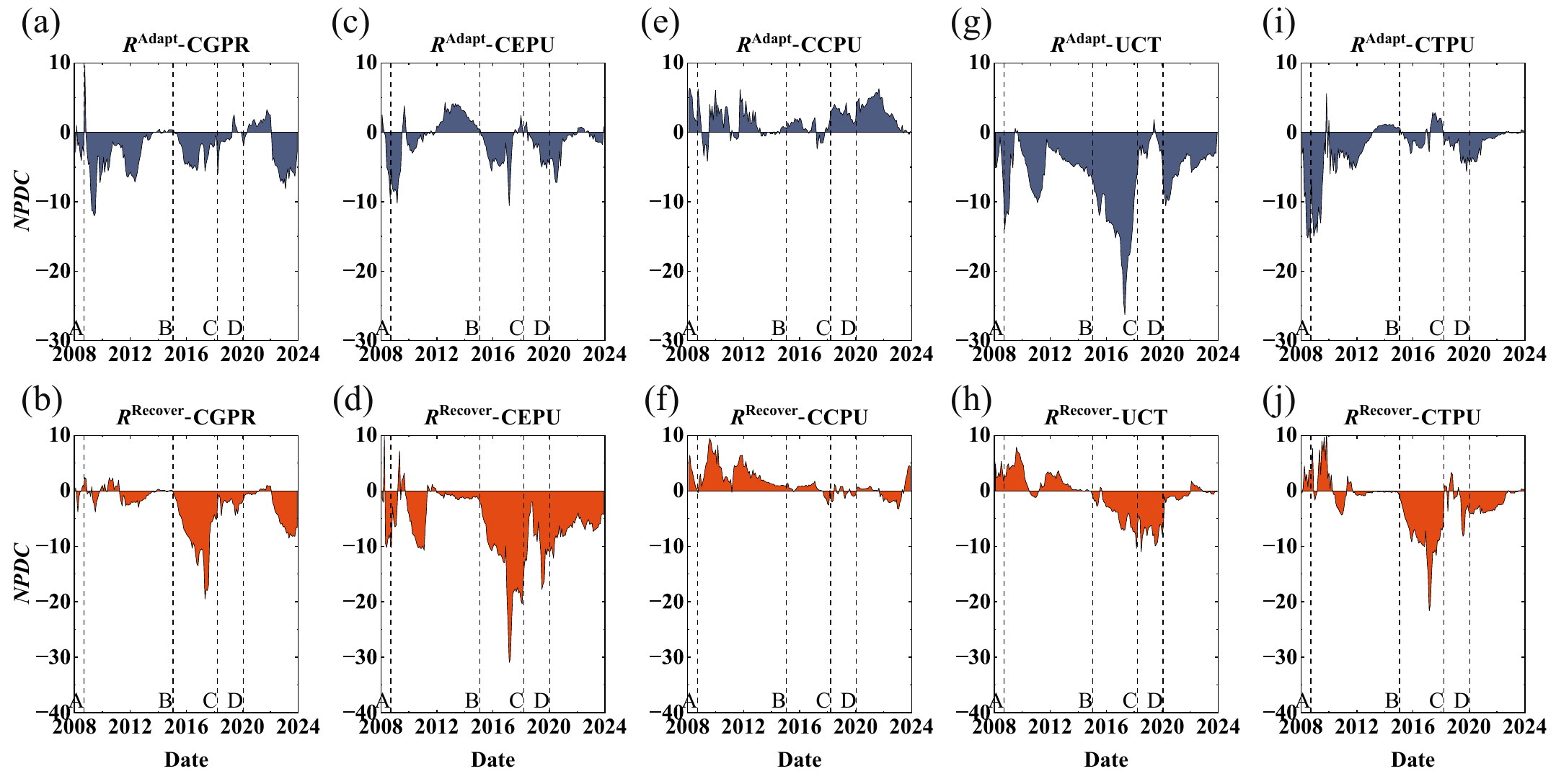}
\caption{Dynamic net pairwise directional connectedness between resilience of China's financial total market and uncertainties. \normalfont{In this figure, the black dotted lines indicate the timing of major events, as shown in Fig.~\ref{Fig_FinResilience_AdaRec}.}}
\label{Fig_NPDC_ShockToTotal}
\end{figure}

Given that the baseline regression results mainly demonstrated the average impact of uncertainty on resilience, we employed the TVP-VAR-DY model to investigate the dynamic impact of five uncertainties on the resilience of China's total financial market, as shown in Fig.~\ref{Fig_NPDC_ShockToTotal}. Notably, the outbreaks of the selected four major events significantly affected changes in volatility spillovers from uncertainties to resilience. For example, the spillovers from CGPR, CEPU, UCT, and CTPU to the adaptability and recoverability of the total market increased markedly during the 2015 stock market crash, indicating that changes in external uncertainties led to pronounced risk resonance in China's financial market, although resilience remained relatively stable during this period. For CCPU, its impact on resilience was not noticeable, consistent with the regression results. Importantly, we find that the US-Sino tension index generated more pronounced spillovers to the adaptability of China's financial market, highlighting that maintaining a stable relationship between the two countries is important for a resilient domestic financial market.

To examine whether external uncertainties had heterogeneous effects on the resilience of different financial sub-markets, we report the baseline regression results in Table~\ref{Table_Regression_Submarket}. The findings indicate that the same external shock exerted varying influences across markets. Specifically, CGPR, CEPU, and UCT significantly reduced the resilience of the stock and bulk commodity markets, whereas their effects on the monetary, bond, and foreign exchange markets were not uniformly significant. These results suggest that the resilience of the stock and bulk commodity markets was more sensitive to macroeconomic and international uncertainties, as capital in these markets tended to fluctuate more sharply in response to elevated risk levels \citep{FJ-Kannadhasan-Das-2020-FinancResLett}. In contrast, the resilience of the foreign exchange market was largely unaffected by CGPR and CEPU, suggesting that the influence of these uncertainties is transmitted to other markets through indirect channels rather than directly through the more tightly regulated foreign exchange market.

For CCPU, its impact on the resilience of China's total market was not significant, while it exerted positive effects on the resilience of the bond and foreign exchange markets. On one hand, as previous studies suggest, green bonds can hedge risks arising from climate policy uncertainty during specific periods \citep{FJ-Wu-Li-Su-2024-EconAnalPolicy,FJ-Naifar-2024-FinancResLett,FJ-Zhang-Shi-Zhao-Yang-2025-RiskManag,FJ-Ren-Li-Ji-Zhai-2025-ApplEconLett}, thereby supporting a more resilient bond market. On the other hand, heightened climate policy uncertainty may lead to a reallocation of capital flows and increased safe-haven demand, enabling the foreign exchange market to better manage volatility and absorb risks \citep{FJ-Ayadi-BenOmrane-Panah-2025-IntRevFinancAnal}.

\begin{table}[htb!]
\centering
\caption{Regression results of external shocks on resilience of China's financial sub-markets}
\label{Table_Regression_Submarket}
\footnotesize
\setlength{\tabcolsep}{4pt}
\renewcommand{\tabcolsep}{0.8mm}
\begin{tabular}{l*{10}{c}}
\toprule
& \multicolumn{5}{c}{$R^\text{Adapt}$} & \multicolumn{5}{c}{$R^\text{Recover}$} \\
\cmidrule(lr){2-6} \cmidrule(lr){7-11}
& Monetary & Stock & Bond & Exchange & Bulk & Monetary & Stock & Bond & Exchange & Bulk \\
\midrule
CGPR & $-0.375^{***}$ & $-0.198^{***}$ & $0.046$ & $0.078$ & $-0.317^{***}$ & $-0.154^{**}$ & $-0.338^{***}$ & $0.170^{**}$ & $-0.024$ & $-0.303^{***}$ \\
     & $(-5.410)$ & $(-2.843)$ & $(0.636)$ & $(1.067)$ & $(-4.586)$ & $(-2.337)$ & $(-5.344)$ & $(2.436)$ & $(-0.351)$ & $(-4.471)$ \\
CEPU & $-0.571^{***}$ & $-0.251^{***}$ & $0.129^{*}$ & $-0.083$ & $-0.179^{**}$ & $-0.187^{***}$ & $-0.428^{***}$ & $0.249^{***}$ & $-0.077$ & $-0.231^{***}$ \\
     & $(-8.353)$ & $(-3.372)$ & $(1.677)$ & $(-1.055)$ & $(-2.311)$ & $(-2.642)$ & $(-6.474)$ & $(3.357)$ & $(-1.041)$ & $(-3.077)$ \\
CCPU & $-0.130^{*}$ & $0.082$ & $0.251^{***}$ & $0.246^{***}$ & $-0.312^{***}$ & $0.062$ & $-0.168^{**}$ & $0.266^{***}$ & $0.140^{**}$ & $0.112$ \\
     & $(-1.721)$ & $(1.131)$ & $(3.524)$ & $(3.386)$ & $(-4.385)$ & $(0.911)$ & $(-2.447)$ & $(3.809)$ & $(2.018)$ & $(-1.540)$ \\
UCT  & $-0.390^{***}$ & $-0.241^{***}$ & $0.086$ & $-0.059$ & $-0.182^{**}$ & $-0.118$ & $-0.352^{***}$ & $0.198^{**}$ & $-0.195^{***}$ & $-0.231^{***}$ \\
     & $(-5.019)$ & $(-3.129)$ & $(1.074)$ & $(-0.720)$ & $(-2.269)$ & $(-1.598)$ & $(-4.964)$ & $(2.552)$ & $(-2.604)$ & $(-2.982)$ \\
CTPU & $-0.451^{***}$ & $-0.136$ & $-0.067$ & $-0.149^{*}$ & $-0.236^{***}$ & $-0.115$ & $-0.246^{***}$ & $0.171^{**}$ & $-0.110$ & $-0.239^{***}$ \\
     & $(-5.556)$ & $(-1.636)$ & $(0.795)$ & $(-1.741)$ & $(-2.801)$ & $(-1.465)$ & $(-3.160)$ & $(2.073)$ & $(-1.366)$ & $(-2.906)$ \\
Control & Yes & Yes & Yes & Yes & Yes & Yes & Yes & Yes & Yes & Yes \\
\bottomrule
\end{tabular}
\begin{flushleft}
\footnotesize
Notes: Values in parentheses represent t-values. The asterisks $^{*}$, $^{**}$, and $^{***}$ indicate significance at the 10\%, 5\%, and 1\% level, respectively.
\end{flushleft}
\end{table}

\begin{figure}[htb!]
\centering
\includegraphics[width=1\textwidth]{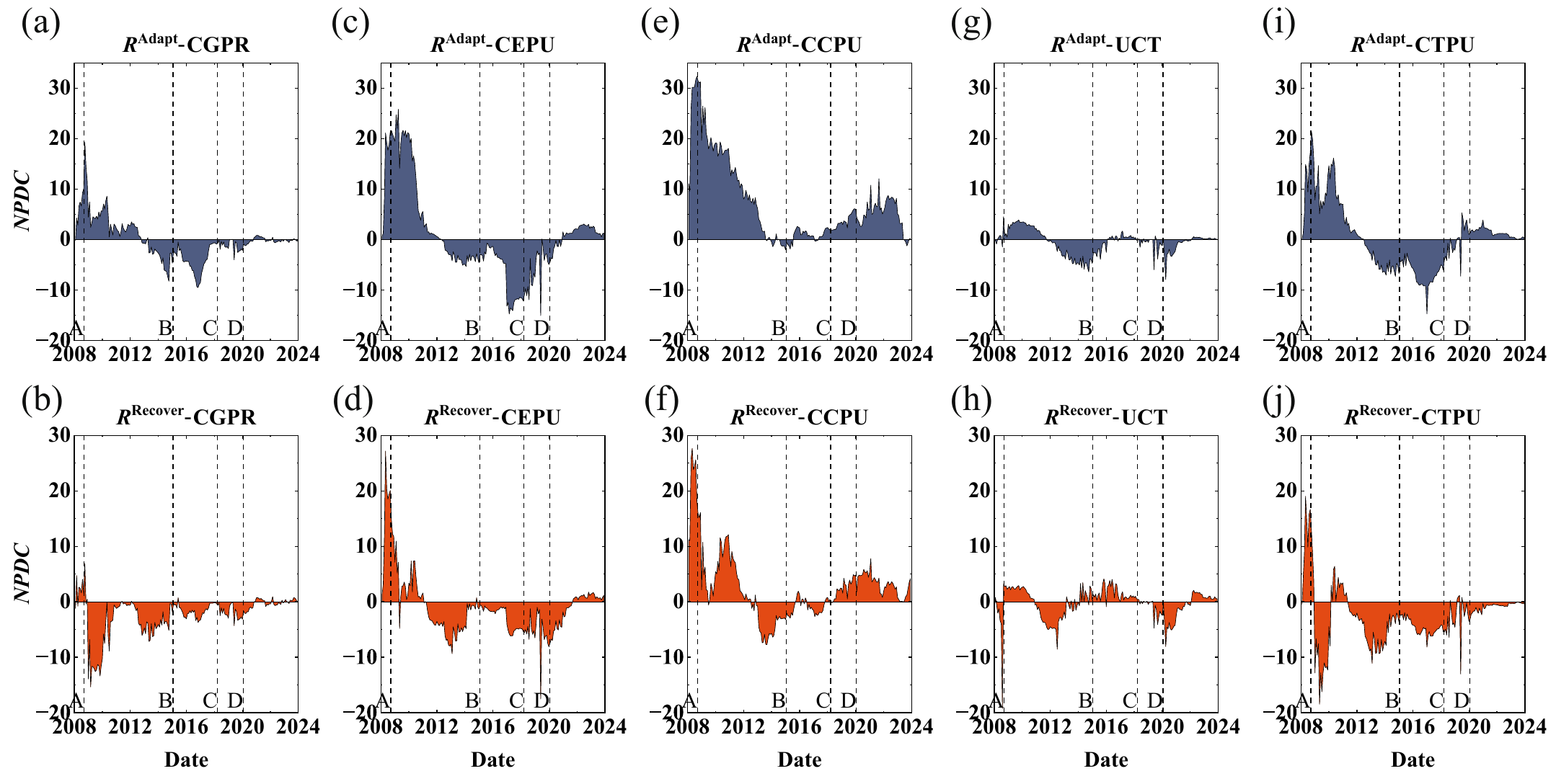}
\caption{Dynamic net pairwise directional connectedness between resilience of China's stock market and uncertainties. \normalfont{In this figure, the black dotted lines indicate the timing of major events, as shown in Fig.~\ref{Fig_FinResilience_AdaRec}.}}
\label{Fig_NPDC_ShockToStock}
\end{figure}

The regression results provide evidence of the average impact of uncertainty on resilience for sub-markets, while it may not be stable over time. Similarly, we employ the TVP-VAR-DY model to investigate the dynamic impact of five uncertainties on the resilience of China's financial sub-markets. Limited by space, we focus on the stock market\textemdash the most representative component of the financial market\textemdash for further analysis. Fig.~\ref{Fig_NPDC_ShockToStock} depicts the dynamic net pairwise directional connectedness between uncertainties and the resilience of the stock market. During the 2008 global financial crisis, we find that the resilience of China's stock market contributed more spillovers to its uncertainties, particularly to CEPU, CCPU, and CTPU, indicating that China's uncertainties during this period were correlated with changes in the stock market's ability to resist and recover from risk \citep{FJ-Uddin-Chowdhury-Anderson-Chaudhuri-2021-JBusRes}.

Notably, CGPR and CTPU mainly exhibited net spillovers to the recoverability of the stock market, indicating greater sensitivity of recoverability to the volatility of geopolitical and trade policy risks. Compared to CEPU, changes in CGPR and CTPU were more abrupt and had the potential to trigger systemic shocks to domestic supply chains and trade order \citep{FJ-Gaies-2025-EconLett,FJ-Alqahtani-Klein-2021-Energy}. Such disruptions could significantly undermine market confidence, which serves as the fundamental foundation for the resilience and recovery of the stock market \citep{FJ-Konzelmann-Wilkinson-FovargueDavies-Sankey-2010-CambrJEcon}.

\section{Conclusion}
\label{Sec_Conclusion}

In an era of escalating uncertainty, understanding and enhancing financial market resilience is paramount for economic stability, particularly for a major emerging economy like China \citep{FJ-Abiad-Bluedorn-Guajardo-Topalova-2015-WorldDev}. This paper addresses this challenge by employing an adaptability- and recoverability-based framework to measure the resilience of China's financial market and systematically investigating the impact of various uncertainties on its dynamics.

Our primary contribution is the construction of dynamic indices for adaptability and recoverability, which reveal that China's financial resilience is not static but a highly event-driven process. These indicators effectively capture sharp deteriorations during major external crises, such as the 2008 global financial crisis and the COVID-19 pandemic, thereby validating their efficacy. Crucially, resilience is not monolithic, where we find that shocks to the resilience of one sub-market rapidly propagate throughout the system. The DY connectedness analysis identifies a clear structure in this contagion: the foreign exchange market consistently emerges as the primary systemic transmitter of resilience shocks, while the more sentiment-driven stock and bulk commodity markets often act as net recipients of spillovers. This underscores that channels of financial contagion are as important as the initial shock itself.

Building on this, we find that related uncertainties act as potent suppressors of China's financial market resilience. Geopolitical risks, economic policy uncertainty, and the U.S.-China tension exert a robust negative influence on both the market's capacity to adapt to shocks and to recover from them. However, the impact is not uniform across all forms of uncertainty. Climate policy uncertainty, in contrast, exhibits a more nuanced, sometimes positive, relationship, particularly enhancing the resilience of the bond and foreign exchange markets. This suggests that as capital reallocates in response to long-term climate risks, it may paradoxically bolster the stability of certain safe-haven assets.

The policy implications of our findings are substantial. First, regulators should adopt a systemic rather than siloed perspective on financial stability. The high degree of connectedness in resilience indicates that a weakness in one market can rapidly become a system-wide vulnerability. In particular, monitoring the foreign exchange market can serve as a crucial early warning system for systemic risk transmission. Second, the heterogeneous impact of different uncertainties calls for nuanced and targeted policy responses. A one-size-fits-all approach to mitigating uncertainty is likely to be suboptimal. Policies aimed at improving market confidence and providing clear forward guidance on economic and trade policy are critical for building buffers against shocks. Finally, our results suggest that while navigating the green transition may introduce new uncertainties, it can also create opportunities to strengthen the resilience of specific market segments, such as the green bond market.

This study is not without limitations. Our resilience metric, predicated on responses to a global external shock (VIX), may under-identify resilience to domestically generated structural shocks, such as the 2015 stock market crash. Future research could develop a more comprehensive resilience framework that incorporates both external and internal shock sources. Also, our measure focuses on the magnitude of the response, not its direction. By using the absolute value of the IRF, we treat both positive and negative deviations from the original state as a loss of stability. As some studies note \citep{FJ-Lyu-Wei-Hu-Yang-2021-Energy,FJ-BenSaida-2019-JFinancMark}, future research could categorise it as either a good or a bad response, further providing a better framework of resilience measurement. Additionally, exploring the feedback loop—how diminished financial resilience might itself amplify economic and policy uncertainty—presents a fruitful avenue for further inquiry.

In conclusion, this paper demonstrates that China's financial market resilience is a dynamic, interconnected system that is highly sensitive to the prevailing uncertainty landscape. As China continues to navigate a complex global environment, building robust defenses against geopolitical and economic policy shocks is not merely a matter of financial market stability but a prerequisite for sustainable economic development.

\section*{Acknowledgment}

This work was partly supported by the National Natural Science Foundation of China (72171083) and the Fundamental Research Funds for the Central Universities.

\section*{Declaration of competing interest}

The authors declare that they have no known competing financial interests or personal relationships that could have appeared to influence the work reported in this paper.

\section*{Data availability}

Data will be made available on request.


\bibliographystyle{elsarticle-harv}
\bibliography{Bib1,Bib2,BibCHN,BibITN,BibRCE,BibRobustNet}

\begin{thebibliography}{90}
\expandafter\ifx\csname natexlab\endcsname\relax\def\natexlab#1{#1}\fi
\providecommand{\url}[1]{\texttt{#1}}
\providecommand{\href}[2]{#2}
\providecommand{\path}[1]{#1}
\providecommand{\DOIprefix}{doi:}
\providecommand{\ArXivprefix}{arXiv:}
\providecommand{\URLprefix}{URL: }
\providecommand{\Pubmedprefix}{pmid:}
\providecommand{\doi}[1]{\href{http://dx.doi.org/#1}{\path{#1}}}
\providecommand{\Pubmed}[1]{\href{pmid:#1}{\path{#1}}}
\providecommand{\bibinfo}[2]{#2}
\ifx\xfnm\relax \def\xfnm[#1]{\unskip,\space#1}\fi
\bibitem[{Abiad et~al.(2015)Abiad, Bluedorn, Guajardo and
  Topalova}]{FJ-Abiad-Bluedorn-Guajardo-Topalova-2015-WorldDev}
\bibinfo{author}{Abiad, A.}, \bibinfo{author}{Bluedorn, J.},
  \bibinfo{author}{Guajardo, J.}, \bibinfo{author}{Topalova, P.},
  \bibinfo{year}{2015}.
\newblock \bibinfo{title}{The rising resilience of emerging market and
  developing economies}.
\newblock \bibinfo{journal}{World Development} \bibinfo{volume}{72},
  \bibinfo{pages}{1--26}.
\newblock \DOIprefix\doi{10.1016/j.worlddev.2015.02.005}.
\bibitem[{Alqahtani and Klein(2021)}]{FJ-Alqahtani-Klein-2021-Energy}
\bibinfo{author}{Alqahtani, A.}, \bibinfo{author}{Klein, T.},
  \bibinfo{year}{2021}.
\newblock \bibinfo{title}{Oil price changes, uncertainty, and geopolitical
  risks: on the resilience of {GCC} countries to global tensions}.
\newblock \bibinfo{journal}{Energy} \bibinfo{volume}{236},
  \bibinfo{pages}{121541}.
\newblock \DOIprefix\doi{10.1016/j.energy.2021.121541}.
\bibitem[{Antonakakis et~al.(2021)Antonakakis, Chatziantoniou and
  Gabauer}]{FJ-Antonakakis-Chatziantoniou-Gabauer-2021-IntJFinancEcon}
\bibinfo{author}{Antonakakis, N.}, \bibinfo{author}{Chatziantoniou, I.},
  \bibinfo{author}{Gabauer, D.}, \bibinfo{year}{2021}.
\newblock \bibinfo{title}{{The impact of Euro through time: Exchange rate
  dynamics under different regimes}}.
\newblock \bibinfo{journal}{International Journal of Finance \& Economics}
  \bibinfo{volume}{26}, \bibinfo{pages}{1375--1408}.
\newblock \DOIprefix\doi{10.1002/ijfe.1854}.
\bibitem[{Ayadi et~al.(2025)Ayadi, Ben~Omrane and
  Panah}]{FJ-Ayadi-BenOmrane-Panah-2025-IntRevFinancAnal}
\bibinfo{author}{Ayadi, M.A.}, \bibinfo{author}{Ben~Omrane, W.},
  \bibinfo{author}{Panah, P.G.}, \bibinfo{year}{2025}.
\newblock \bibinfo{title}{Foreign exchange markets, climate risks and
  contextual news: an intraday analysis}.
\newblock \bibinfo{journal}{International Review of Financial Analysis}
  \bibinfo{volume}{102}, \bibinfo{pages}{104103}.
\newblock \DOIprefix\doi{10.1016/j.irfa.2025.104103}.
\bibitem[{Battiston et~al.(2021)Battiston, Dafermos and
  Monasterolo}]{FJ-Battiston-Dafermos-Monasterolo-2021-JFinancStab}
\bibinfo{author}{Battiston, S.}, \bibinfo{author}{Dafermos, Y.},
  \bibinfo{author}{Monasterolo, I.}, \bibinfo{year}{2021}.
\newblock \bibinfo{title}{Climate risks and financial stability}.
\newblock \bibinfo{journal}{Journal of Financial Stability}
  \bibinfo{volume}{54}, \bibinfo{pages}{100867}.
\newblock \DOIprefix\doi{10.1016/j.jfs.2021.100867}.
\bibitem[{Battiston et~al.(2012)Battiston, Gatti, Gallegati, Greenwald and
  Stiglitz}]{FJ-Battiston-Gatti-Gallegati-Greenwald-Stiglitz-2012-JEconDynControl}
\bibinfo{author}{Battiston, S.}, \bibinfo{author}{Gatti, D.D.},
  \bibinfo{author}{Gallegati, M.}, \bibinfo{author}{Greenwald, B.},
  \bibinfo{author}{Stiglitz, J.E.}, \bibinfo{year}{2012}.
\newblock \bibinfo{title}{Liaisons dangereuses: increasing connectivity, risk
  sharing, and systemic risk}.
\newblock \bibinfo{journal}{Journal of Economic Dynamics \& Control}
  \bibinfo{volume}{36}, \bibinfo{pages}{1121--1141}.
\newblock \DOIprefix\doi{10.1016/j.jedc.2012.04.001}.
\bibitem[{BenSa\"{i}da(2019)}]{FJ-BenSaida-2019-JFinancMark}
\bibinfo{author}{BenSa\"{i}da, A.}, \bibinfo{year}{2019}.
\newblock \bibinfo{title}{Good and bad volatility spillovers: an asymmetric
  connectedness}.
\newblock \bibinfo{journal}{Journal of Financial Markets} \bibinfo{volume}{43},
  \bibinfo{pages}{78--95}.
\newblock \DOIprefix\doi{10.1016/j.finmar.2018.12.005}.
\bibitem[{Berestycki et~al.(2022)Berestycki, Carattini, Dechezlepr{\^e}tre and
  Kruse}]{FJ-Berestycki-Carattini-DechezleprEtre-Kruse-2022-OECD}
\bibinfo{author}{Berestycki, C.}, \bibinfo{author}{Carattini, S.},
  \bibinfo{author}{Dechezlepr{\^e}tre, A.}, \bibinfo{author}{Kruse, T.},
  \bibinfo{year}{2022}.
\newblock \bibinfo{title}{Measuring and assessing the effects of climate policy
  uncertainty}.
\newblock \DOIprefix\doi{10.1787/34483d83-en}. \bibinfo{note}{{OECD Economics
  Department Working Paper}}.
\bibitem[{Blot et~al.(2015)Blot, Creel, Hubert, Labondance and
  Saraceno}]{FJ-Blot-Creel-Hubert-Labondance-Saraceno-2015-JFinancStab}
\bibinfo{author}{Blot, C.}, \bibinfo{author}{Creel, J.},
  \bibinfo{author}{Hubert, P.}, \bibinfo{author}{Labondance, F.},
  \bibinfo{author}{Saraceno, F.}, \bibinfo{year}{2015}.
\newblock \bibinfo{title}{Assessing the link between price and financial
  stability}.
\newblock \bibinfo{journal}{Journal of Financial Stability}
  \bibinfo{volume}{16}, \bibinfo{pages}{71--88}.
\newblock \DOIprefix\doi{10.1016/j.jfs.2014.12.003}.
\bibitem[{Bruneau et~al.(2003)Bruneau, Chang, Eguchi, Lee, O'Rourke, Reinhorn,
  Shinozuka, Tierney, Wallace and
  Von~Winterfeldt}]{FJ-Bruneau-Chang-Eguchi-Lee-ORourke-Reinhorn-Shinozuka-Tierney-Wallace-VonWinterfeldt-2003-EarthqSpectra}
\bibinfo{author}{Bruneau, M.}, \bibinfo{author}{Chang, S.E.},
  \bibinfo{author}{Eguchi, R.T.}, \bibinfo{author}{Lee, G.C.},
  \bibinfo{author}{O'Rourke, T.D.}, \bibinfo{author}{Reinhorn, A.M.},
  \bibinfo{author}{Shinozuka, M.}, \bibinfo{author}{Tierney, K.},
  \bibinfo{author}{Wallace, W.A.}, \bibinfo{author}{Von~Winterfeldt, D.},
  \bibinfo{year}{2003}.
\newblock \bibinfo{title}{A framework to quantitatively assess and enhance the
  seismic resilience of communities}.
\newblock \bibinfo{journal}{Earthquake Spectra} \bibinfo{volume}{19},
  \bibinfo{pages}{733--752}.
\newblock \DOIprefix\doi{10.1193/1.1623497}.
\bibitem[{Brunnermeier(2024)}]{FJ-Brunnermeier-2024-JFinanc}
\bibinfo{author}{Brunnermeier, M.K.}, \bibinfo{year}{2024}.
\newblock \bibinfo{title}{Presidential address: Macrofinance and resilience}.
\newblock \bibinfo{journal}{Journal of Finance} \bibinfo{volume}{79},
  \bibinfo{pages}{3683--3728}.
\newblock \DOIprefix\doi{10.1111/jofi.13403}.
\bibitem[{Bui et~al.(2017)Bui, Scheule and
  Wu}]{FJ-Bui-Scheule-Wu-2017-JFinancStab}
\bibinfo{author}{Bui, C.}, \bibinfo{author}{Scheule, H.}, \bibinfo{author}{Wu,
  E.}, \bibinfo{year}{2017}.
\newblock \bibinfo{title}{The value of bank capital buffers in maintaining
  financial system resilience}.
\newblock \bibinfo{journal}{Journal of Financial Stability}
  \bibinfo{volume}{33}, \bibinfo{pages}{23--40}.
\newblock \DOIprefix\doi{10.1016/j.jfs.2017.10.006}.
\bibitem[{Cevik et~al.(2024)Cevik, Terzioglu, Kilic, Bugan and
  Dibooglu}]{FJ-Cevik-Terzioglu-Kilic-Bugan-Dibooglu-2024-ResIntBusFinanc}
\bibinfo{author}{Cevik, E.I.}, \bibinfo{author}{Terzioglu, H.C.},
  \bibinfo{author}{Kilic, Y.}, \bibinfo{author}{Bugan, M.F.},
  \bibinfo{author}{Dibooglu, S.}, \bibinfo{year}{2024}.
\newblock \bibinfo{title}{Interconnectedness and systemic risk: Evidence from
  global stock markets}.
\newblock \bibinfo{journal}{Research in International Business and Finance}
  \bibinfo{volume}{69}, \bibinfo{pages}{102282}.
\newblock \DOIprefix\doi{10.1016/j.ribaf.2024.102282}.
\bibitem[{Cheema et~al.(2022)Cheema, Faff and
  Szulczyk}]{FJ-Cheema-Faff-Szulczyk-2022-IntRevFinancAnal}
\bibinfo{author}{Cheema, M.A.}, \bibinfo{author}{Faff, R.},
  \bibinfo{author}{Szulczyk, K.R.}, \bibinfo{year}{2022}.
\newblock \bibinfo{title}{The 2008 global financial crisis and {COVID}-19
  pandemic: how safe are the safe haven assets?}
\newblock \bibinfo{journal}{International Review of Financial Analysis}
  \bibinfo{volume}{83}, \bibinfo{pages}{102316}.
\newblock \DOIprefix\doi{10.1016/j.irfa.2022.102316}.
\bibitem[{Chen and Sun(2022)}]{FJ-Chen-Sun-2022-NAmEconFinanc}
\bibinfo{author}{Chen, B.X.}, \bibinfo{author}{Sun, Y.L.},
  \bibinfo{year}{2022}.
\newblock \bibinfo{title}{The impact of vix on {C}hina's financial market: a
  new perspective based on high-dimensional and time-varying methods}.
\newblock \bibinfo{journal}{North American Journal of Economics and Finance}
  \bibinfo{volume}{63}, \bibinfo{pages}{101831}.
\newblock \DOIprefix\doi{10.1016/j.najef.2022.101831}.
\bibitem[{Chen and Gong(2019)}]{FJ-Chen-Gong-2019-IntRevEconFinanc}
\bibinfo{author}{Chen, Q.}, \bibinfo{author}{Gong, Y.}, \bibinfo{year}{2019}.
\newblock \bibinfo{title}{The economic sources of {C}hina's csi 300 spot and
  futures volatilities before and after the 2015 stock market crisis}.
\newblock \bibinfo{journal}{International Review of Economics \& Finance}
  \bibinfo{volume}{64}, \bibinfo{pages}{102--121}.
\newblock \DOIprefix\doi{10.1016/j.iref.2019.05.017}.
\bibitem[{Chen and He(2022)}]{FJ-Chen-He-2022-Sustainability}
\bibinfo{author}{Chen, X.}, \bibinfo{author}{He, Y.}, \bibinfo{year}{2022}.
\newblock \bibinfo{title}{The impact of financial resilience and steady growth
  on high-quality economic development-based on a heterogeneous intermediary
  effect analysis}.
\newblock \bibinfo{journal}{Sustainability} \bibinfo{volume}{14},
  \bibinfo{pages}{14748}.
\newblock \DOIprefix\doi{10.3390/su142214748}.
\bibitem[{Chen et~al.(2024)Chen, Ma, Chen and
  Yang}]{FJ-Chen-Ma-Chen-Yang-2024-TranspResPartD}
\bibinfo{author}{Chen, X.}, \bibinfo{author}{Ma, S.}, \bibinfo{author}{Chen,
  L.}, \bibinfo{author}{Yang, L.}, \bibinfo{year}{2024}.
\newblock \bibinfo{title}{Resilience measurement and analysis of intercity
  public transportation network}.
\newblock \bibinfo{journal}{Transportation Research Part D}
  \bibinfo{volume}{131}, \bibinfo{pages}{104202}.
\newblock \DOIprefix\doi{10.1016/j.trd.2024.104202}.
\bibitem[{Chen and Sun(2024)}]{FJ-Chen-Sun-2024-EconLett}
\bibinfo{author}{Chen, Y.}, \bibinfo{author}{Sun, C.}, \bibinfo{year}{2024}.
\newblock \bibinfo{title}{A new method for measuring financial resilience}.
\newblock \bibinfo{journal}{Economics Letters} \bibinfo{volume}{242},
  \bibinfo{pages}{111883}.
\newblock \DOIprefix\doi{10.1016/j.econlet.2024.111883}.
\bibitem[{Chen et~al.(2025)Chen, Sun and
  Zhang}]{FJ-Chen-Sun-Zhang-2025-JFinancStab}
\bibinfo{author}{Chen, Y.}, \bibinfo{author}{Sun, C.}, \bibinfo{author}{Zhang,
  X.}, \bibinfo{year}{2025}.
\newblock \bibinfo{title}{Analyzing and forecasting {C}hina's financial
  resilience: measurement techniques and identification of key influencing
  factors}.
\newblock \bibinfo{journal}{Journal of Financial Stability}
  \bibinfo{volume}{76}, \bibinfo{pages}{101372}.
\newblock \DOIprefix\doi{10.1016/j.jfs.2025.101372}.
\bibitem[{Chen et~al.(2023)Chen, Mo, Qin and
  Yang}]{FJ-Chen-Mo-Qin-Yang-2023-Pac-BasinFinancJ}
\bibinfo{author}{Chen, Y.L.}, \bibinfo{author}{Mo, W.S.}, \bibinfo{author}{Qin,
  R.L.}, \bibinfo{author}{Yang, J.J.}, \bibinfo{year}{2023}.
\newblock \bibinfo{title}{Return spillover across {C}hina's financial markets}.
\newblock \bibinfo{journal}{Pacific-Basin Finance Journal}
  \bibinfo{volume}{80}, \bibinfo{pages}{102057}.
\newblock \DOIprefix\doi{10.1016/j.pacfin.2023.102057}.
\bibitem[{Chen et~al.(2020)Chen, He and Liu}]{FJ-Chen-He-Liu-2020-JFinancEcon}
\bibinfo{author}{Chen, Z.}, \bibinfo{author}{He, Z.}, \bibinfo{author}{Liu,
  C.}, \bibinfo{year}{2020}.
\newblock \bibinfo{title}{The financing of local government in {C}hina:
  stimulus loan wanes and shadow banking waxes}.
\newblock \bibinfo{journal}{Journal of Financial Economics}
  \bibinfo{volume}{137}, \bibinfo{pages}{42--71}.
\newblock \DOIprefix\doi{10.1016/j.jfineco.2019.07.009}.
\bibitem[{Chernykh et~al.(2023)Chernykh, Davydov and
  Sihvonen}]{FJ-Chernykh-Davydov-Sihvonen-2023-JFinancStab}
\bibinfo{author}{Chernykh, L.}, \bibinfo{author}{Davydov, D.},
  \bibinfo{author}{Sihvonen, J.}, \bibinfo{year}{2023}.
\newblock \bibinfo{title}{Financial stability and public confidence in banks}.
\newblock \bibinfo{journal}{Journal of Financial Stability}
  \bibinfo{volume}{69}, \bibinfo{pages}{101187}.
\newblock \DOIprefix\doi{10.1016/j.jfs.2023.101187}.
\bibitem[{Ciullo et~al.(2023)Ciullo, Strobl, Meiler, Martius and
  Bresch}]{FJ-Ciullo-Strobl-Meiler-Martius-Bresch-2023-NatCommun}
\bibinfo{author}{Ciullo, A.}, \bibinfo{author}{Strobl, E.},
  \bibinfo{author}{Meiler, S.}, \bibinfo{author}{Martius, O.},
  \bibinfo{author}{Bresch, D.N.}, \bibinfo{year}{2023}.
\newblock \bibinfo{title}{Increasing countries' financial resilience through
  global catastrophe risk pooling}.
\newblock \bibinfo{journal}{Nature Communications} \bibinfo{volume}{14},
  \bibinfo{pages}{922}.
\newblock \DOIprefix\doi{10.1038/s41467-023-36539-4}.
\bibitem[{Daadmehr(2024)}]{FJ-Daadmehr-2024-RiskManag}
\bibinfo{author}{Daadmehr, E.}, \bibinfo{year}{2024}.
\newblock \bibinfo{title}{Workplace sustainability or financial resilience?
  composite-financial resilience index}.
\newblock \bibinfo{journal}{Risk Management} \bibinfo{volume}{26},
  \bibinfo{pages}{7}.
\newblock \DOIprefix\doi{10.1057/s41283-023-00139-9}.
\bibitem[{Diebold and Yilmaz(2012)}]{FJ-Diebold-Yilmaz-2012-IntJForecast}
\bibinfo{author}{Diebold, F.X.}, \bibinfo{author}{Yilmaz, K.},
  \bibinfo{year}{2012}.
\newblock \bibinfo{title}{Better to give than to receive: {P}redictive
  directional measurement of volatility spillovers}.
\newblock \bibinfo{journal}{International Journal of Forecasting}
  \bibinfo{volume}{28}, \bibinfo{pages}{57--66}.
\newblock \DOIprefix\doi{10.1016/j.ijforecast.2011.02.006}.
\bibitem[{Dornbusch and Fischer(1980)}]{FJ-Dornbusch-Fischer-1980-AmEconRev}
\bibinfo{author}{Dornbusch, R.}, \bibinfo{author}{Fischer, S.},
  \bibinfo{year}{1980}.
\newblock \bibinfo{title}{Exchange rates and the current account}.
\newblock \bibinfo{journal}{American Economic Review} \bibinfo{volume}{70},
  \bibinfo{pages}{960--971}.
\bibitem[{Fang et~al.(2022)Fang, Chu, Fu and
  Fang}]{FJ-Fang-Chu-Fu-Fang-2022-TranspResPartD}
\bibinfo{author}{Fang, C.}, \bibinfo{author}{Chu, Y.}, \bibinfo{author}{Fu,
  H.}, \bibinfo{author}{Fang, Y.}, \bibinfo{year}{2022}.
\newblock \bibinfo{title}{On the resilience assessment of complementary
  transportation networks under natural hazards}.
\newblock \bibinfo{journal}{Transportation Research Part D}
  \bibinfo{volume}{109}, \bibinfo{pages}{103331}.
\newblock \DOIprefix\doi{10.1016/j.trd.2022.103331}.
\bibitem[{Folke(2006)}]{FJ-Folke-2006-GlobEnvironChange}
\bibinfo{author}{Folke, C.}, \bibinfo{year}{2006}.
\newblock \bibinfo{title}{Resilience: the emergence of a perspective for
  social-ecological systems analyses}.
\newblock \bibinfo{journal}{Global Environmental Change} \bibinfo{volume}{16},
  \bibinfo{pages}{253--267}.
\newblock \DOIprefix\doi{10.1016/j.gloenvcha.2006.04.002}.
\bibitem[{Forbes and Rigobon(2002)}]{FJ-Forbes-Rigobon-2002-JFinanc}
\bibinfo{author}{Forbes, K.}, \bibinfo{author}{Rigobon, R.},
  \bibinfo{year}{2002}.
\newblock \bibinfo{title}{No contagion, only interdependence: measuring stock
  market comovements}.
\newblock \bibinfo{journal}{Journal of Finance} \bibinfo{volume}{57},
  \bibinfo{pages}{2223--2261}.
\newblock \DOIprefix\doi{10.1111/0022-1082.00494}.
\bibitem[{Gagnon and
  Sarsenbayev(2023)}]{FJ-Gagnon-Sarsenbayev-2023-JIntMoneyFinan}
\bibinfo{author}{Gagnon, J.E.}, \bibinfo{author}{Sarsenbayev, M.},
  \bibinfo{year}{2023}.
\newblock \bibinfo{title}{Dollar not so dominant: dollar invoicing has only a
  small effect on trade prices}.
\newblock \bibinfo{journal}{Journal of International Money and Finance}
  \bibinfo{volume}{137}, \bibinfo{pages}{102889}.
\newblock \DOIprefix\doi{10.1016/j.jimonfin.2023.102889}.
\bibitem[{Gaies(2025)}]{FJ-Gaies-2025-EconLett}
\bibinfo{author}{Gaies, B.}, \bibinfo{year}{2025}.
\newblock \bibinfo{title}{Asymmetric effects of tariffs on stock prices:
  Evidence from the us}.
\newblock \bibinfo{journal}{Economics Letters} \bibinfo{volume}{255},
  \bibinfo{pages}{112527}.
\newblock \DOIprefix\doi{10.1016/j.econlet.2025.112527}.
\bibitem[{Geweke(1992)}]{FJ-Geweke-1992-BayesStat}
\bibinfo{author}{Geweke, J.}, \bibinfo{year}{1992}.
\newblock \bibinfo{title}{Evaluating the accuracy of sampling-based approaches
  to the calculations of posterior moments}.
\newblock \bibinfo{journal}{Bayesian Statistics} \bibinfo{volume}{4},
  \bibinfo{pages}{641--649}.
\newblock \URLprefix \url{https://ideas.repec.org/p/fip/fedmsr/148.html}.
\bibitem[{Ghadge et~al.(2022)Ghadge, Er, Ivanov and
  Chaudhuri}]{FJ-Ghadge-Er-Ivanov-Chaudhuri-2022-IntJProdRes}
\bibinfo{author}{Ghadge, A.}, \bibinfo{author}{Er, M.},
  \bibinfo{author}{Ivanov, D.}, \bibinfo{author}{Chaudhuri, A.},
  \bibinfo{year}{2022}.
\newblock \bibinfo{title}{Visualisation of ripple effect in supply chains under
  long-term, simultaneous disruptions: A system dynamics approach}.
\newblock \bibinfo{journal}{International Journal of Production Research}
  \bibinfo{volume}{60}, \bibinfo{pages}{6173--6186}.
\newblock \DOIprefix\doi{10.1080/00207543.2021.1987547}.
\bibitem[{Glebocki and
  Saha(2024)}]{FJ-Glebocki-Saha-2024-JIntFinancMarkInstMoney}
\bibinfo{author}{Glebocki, H.}, \bibinfo{author}{Saha, S.},
  \bibinfo{year}{2024}.
\newblock \bibinfo{title}{Global uncertainty and exchange rate conditions:
  assessing the impact of uncertainty shocks in emerging markets and advanced
  economies}.
\newblock \bibinfo{journal}{Journal of International Financial Markets
  Institutions \& Money} \bibinfo{volume}{96}, \bibinfo{pages}{102060}.
\newblock \DOIprefix\doi{10.1016/j.intfin.2024.102060}.
\bibitem[{Greenwood-Nimmo and
  Tarassow(2016)}]{FJ-GreenwoodNimmo-Tarassow-2016-EconModel}
\bibinfo{author}{Greenwood-Nimmo, M.}, \bibinfo{author}{Tarassow, A.},
  \bibinfo{year}{2016}.
\newblock \bibinfo{title}{Monetary shocks, macroprudential shocks and financial
  stability}.
\newblock \bibinfo{journal}{Economic Modelling} \bibinfo{volume}{56},
  \bibinfo{pages}{11--24}.
\newblock \DOIprefix\doi{10.1016/j.econmod.2016.03.003}.
\bibitem[{Haimes(2009)}]{FJ-Haimes-2009-RiskAnal}
\bibinfo{author}{Haimes, Y.Y.}, \bibinfo{year}{2009}.
\newblock \bibinfo{title}{On the complex definition of risk: A systems-based
  approach}.
\newblock \bibinfo{journal}{Risk Analysis} \bibinfo{volume}{29},
  \bibinfo{pages}{1647--1654}.
\newblock \DOIprefix\doi{10.1111/j.1539-6924.2009.01310.x}.
\bibitem[{Haldane and May(2011)}]{FJ-Haldane-May-2011-Nature}
\bibinfo{author}{Haldane, A.G.}, \bibinfo{author}{May, R.M.},
  \bibinfo{year}{2011}.
\newblock \bibinfo{title}{Systemic risk in banking ecosystems}.
\newblock \bibinfo{journal}{Nature} \bibinfo{volume}{469},
  \bibinfo{pages}{MISSING}.
\newblock \DOIprefix\doi{10.1038/nature09659}.
\bibitem[{Han et~al.(2024)Han, Sun and
  Jiang}]{FJ-Han-Sun-Jiang-2024-FrontEnvironSci}
\bibinfo{author}{Han, J.}, \bibinfo{author}{Sun, Q.}, \bibinfo{author}{Jiang,
  Y.}, \bibinfo{year}{2024}.
\newblock \bibinfo{title}{Studying the risk spillover effects of the carbon
  market and high-carbon-emission industries under economic uncertainty}.
\newblock \bibinfo{journal}{Frontiers in Environmental Science}
  \bibinfo{volume}{12}, \bibinfo{pages}{1407135}.
\newblock \DOIprefix\doi{10.3389/fenvs.2024.1407135}.
\bibitem[{He et~al.(2024)He, Liang and
  Liu}]{FJ-He-Liang-Liu-2024-JIntMoneyFinan}
\bibinfo{author}{He, Q.}, \bibinfo{author}{Liang, B.}, \bibinfo{author}{Liu,
  J.}, \bibinfo{year}{2024}.
\newblock \bibinfo{title}{{RMB internationalization and exchange rate exposure
  of Chinese listed firms}}.
\newblock \bibinfo{journal}{Journal of International Money and Finance}
  \bibinfo{volume}{145}, \bibinfo{pages}{103098}.
\newblock \DOIprefix\doi{10.1016/j.jimonfin.2024.103098}.
\bibitem[{He and Wei(2023)}]{FJ-He-Wei-2023-AnnuRevEcon}
\bibinfo{author}{He, Z.}, \bibinfo{author}{Wei, W.}, \bibinfo{year}{2023}.
\newblock \bibinfo{title}{China's financial system and economy: a review}.
\newblock \bibinfo{journal}{Annual Review of Economics} \bibinfo{volume}{15},
  \bibinfo{pages}{451--483}.
\newblock \DOIprefix\doi{10.1146/annurev-economics-072622-095926}.
\bibitem[{Holling(1973)}]{FJ-Holling-1973-AnnuRevEcolSystemat}
\bibinfo{author}{Holling, C.}, \bibinfo{year}{1973}.
\newblock \bibinfo{title}{Resilience and stability of ecological systems}.
\newblock \bibinfo{journal}{Annual Review of Ecology Evolution and Systematics}
  \bibinfo{volume}{4}, \bibinfo{pages}{1--23}.
\newblock \DOIprefix\doi{10.1146/annurev.es.04.110173.000245}.
\bibitem[{Holl{\'o} et~al.(2012)Holl{\'o}, Kremer and
  Lo~Duca}]{FJ-Hollo-Kremer-LoDuca-2012-ECB}
\bibinfo{author}{Holl{\'o}, D.}, \bibinfo{author}{Kremer, M.},
  \bibinfo{author}{Lo~Duca, M.}, \bibinfo{year}{2012}.
\newblock \bibinfo{title}{{CISS-a composite indicator of systemic stress in the
  financial system}}.
\newblock \bibinfo{note}{ECB Working paper}.
\bibitem[{Hosseini et~al.(2016)Hosseini, Barker and
  Ramirez-Marquez}]{FJ-Hosseini-Barker-RamirezMarquez-2016-ReliabEngSystSaf}
\bibinfo{author}{Hosseini, S.}, \bibinfo{author}{Barker, K.},
  \bibinfo{author}{Ramirez-Marquez, J.E.}, \bibinfo{year}{2016}.
\newblock \bibinfo{title}{A review of definitions and measures of system
  resilience}.
\newblock \bibinfo{journal}{Reliability Engineering \& System Safety}
  \bibinfo{volume}{145}, \bibinfo{pages}{47--61}.
\newblock \DOIprefix\doi{10.1016/j.ress.2015.08.006}.
\bibitem[{Huang et~al.(2014)Huang, Fang and
  Miller}]{FJ-Huang-Fang-Miller-2014-JEmpirFinanc}
\bibinfo{author}{Huang, H.C.R.}, \bibinfo{author}{Fang, W.},
  \bibinfo{author}{Miller, S.M.}, \bibinfo{year}{2014}.
\newblock \bibinfo{title}{Banking market structure, liquidity needs, and
  industrial growth volatility}.
\newblock \bibinfo{journal}{Journal of Empirical Finance} \bibinfo{volume}{26},
  \bibinfo{pages}{1--12}.
\newblock \DOIprefix\doi{10.1016/j.jempfin.2014.01.001}.
\bibitem[{Ivanov(2022)}]{FJ-Ivanov-2022-AnnOperRes}
\bibinfo{author}{Ivanov, D.}, \bibinfo{year}{2022}.
\newblock \bibinfo{title}{Blackout and supply chains: Cross-structural ripple
  effect, performance, resilience and viability impact analysis}.
\newblock \bibinfo{journal}{Annals of Operations Research} ,
  \bibinfo{pages}{1--17}\DOIprefix\doi{10.1007/s10479-022-04754-9}.
\bibitem[{Kannadhasan and Das(2020)}]{FJ-Kannadhasan-Das-2020-FinancResLett}
\bibinfo{author}{Kannadhasan, M.}, \bibinfo{author}{Das, D.},
  \bibinfo{year}{2020}.
\newblock \bibinfo{title}{{Do Asian emerging stock markets react to
  international economic policy uncertainty and geopolitical risk alike? a
  quantile regression approach}}.
\newblock \bibinfo{journal}{Finance Research Letters} \bibinfo{volume}{34},
  \bibinfo{pages}{101276}.
\newblock \DOIprefix\doi{10.1016/j.frl.2019.08.024}.
\bibitem[{Klimek et~al.(2019)Klimek, Poledna and
  Thurner}]{FJ-Klimek-Poledna-Thurner-2019-NatCommun}
\bibinfo{author}{Klimek, P.}, \bibinfo{author}{Poledna, S.},
  \bibinfo{author}{Thurner, S.}, \bibinfo{year}{2019}.
\newblock \bibinfo{title}{Quantifying economic resilience from input--output
  susceptibility to improve predictions of economic growth and recovery}.
\newblock \bibinfo{journal}{Nature Communications} \bibinfo{volume}{10},
  \bibinfo{pages}{1677}.
\newblock \DOIprefix\doi{10.1038/s41467-019-09357-w}.
\bibitem[{Koetter et~al.(2022)Koetter, Krause, Sfrappini and
  Tonzer}]{FJ-Koetter-Krause-Sfrappini-Tonzer-2022-EurEconRev}
\bibinfo{author}{Koetter, M.}, \bibinfo{author}{Krause, T.},
  \bibinfo{author}{Sfrappini, E.}, \bibinfo{author}{Tonzer, L.},
  \bibinfo{year}{2022}.
\newblock \bibinfo{title}{Completing the {E}uropean banking union: capital cost
  consequences for credit providers and corporate borrowers}.
\newblock \bibinfo{journal}{European Economic Review} \bibinfo{volume}{148},
  \bibinfo{pages}{104229}.
\newblock \DOIprefix\doi{10.1016/j.euroecorev.2022.104229}.
\bibitem[{Konzelmann et~al.(2010)Konzelmann, Wilkinson, Fovargue-Davies and
  Sankey}]{FJ-Konzelmann-Wilkinson-FovargueDavies-Sankey-2010-CambrJEcon}
\bibinfo{author}{Konzelmann, S.}, \bibinfo{author}{Wilkinson, F.},
  \bibinfo{author}{Fovargue-Davies, M.}, \bibinfo{author}{Sankey, D.},
  \bibinfo{year}{2010}.
\newblock \bibinfo{title}{Governance, regulation and financial market
  instability: the implications for policy}.
\newblock \bibinfo{journal}{Cambridge Journal of Economics}
  \bibinfo{volume}{34}, \bibinfo{pages}{929--954}.
\newblock \DOIprefix\doi{10.1093/cje/bep086}.
\bibitem[{Kubitza(2025)}]{FJ-Kubitza-2025-JFinancQuantAnal}
\bibinfo{author}{Kubitza, C.}, \bibinfo{year}{2025}.
\newblock \bibinfo{title}{Tackling the volatility paradox: spillover
  persistence and systemic risk}.
\newblock \bibinfo{journal}{Journal of Financial and Quantitative Analysis} ,
  \bibinfo{pages}{MISSING}\DOIprefix\doi{10.1017/S002210902400070X}.
\bibitem[{Li et~al.(2022)Li, Zheng and Liu}]{FJ-Li-Zheng-Liu-2022-RegulGov}
\bibinfo{author}{Li, C.}, \bibinfo{author}{Zheng, H.}, \bibinfo{author}{Liu,
  Y.}, \bibinfo{year}{2022}.
\newblock \bibinfo{title}{The hybrid regulatory regime in turbulent times: the
  role of the state in {C}hina's stock market crisis in 2015-2016}.
\newblock \bibinfo{journal}{Regulation \& Governance} \bibinfo{volume}{16},
  \bibinfo{pages}{392--408}.
\newblock \DOIprefix\doi{10.1111/rego.12340}.
\bibitem[{Liang et~al.(2024)Liang, Goodell and
  Li}]{FJ-Liang-Goodell-Li-2024-JIntFinancMarkInstMoney}
\bibinfo{author}{Liang, C.}, \bibinfo{author}{Goodell, J.W.},
  \bibinfo{author}{Li, X.}, \bibinfo{year}{2024}.
\newblock \bibinfo{title}{Impacts of carbon market and climate policy
  uncertainties on financial and economic stability: Evidence from
  connectedness network analysis}.
\newblock \bibinfo{journal}{Journal of International Financial Markets
  Institutions \& Money} \bibinfo{volume}{92}, \bibinfo{pages}{101977}.
\newblock \DOIprefix\doi{10.1016/j.intfin.2024.101977}.
\bibitem[{Linnemann and Schabert(2015)}]{FJ-Linnemann-Schabert-2015-JIntEcon}
\bibinfo{author}{Linnemann, L.}, \bibinfo{author}{Schabert, A.},
  \bibinfo{year}{2015}.
\newblock \bibinfo{title}{Liquidity premia and interest rate parity}.
\newblock \bibinfo{journal}{Journal of International Economics}
  \bibinfo{volume}{97}, \bibinfo{pages}{178--192}.
\newblock \DOIprefix\doi{10.1016/j.jinteco.2015.03.006}.
\bibitem[{Liu et~al.(2016)Liu, Gu and
  Xing}]{FJ-Liu-Gu-Xing-2016-IntRevEconFinanc}
\bibinfo{author}{Liu, D.}, \bibinfo{author}{Gu, H.}, \bibinfo{author}{Xing,
  T.}, \bibinfo{year}{2016}.
\newblock \bibinfo{title}{The meltdown of the {C}hinese equity market in the
  summer of 2015}.
\newblock \bibinfo{journal}{International Review of Economics \& Finance}
  \bibinfo{volume}{45}, \bibinfo{pages}{504--517}.
\newblock \DOIprefix\doi{10.1016/j.iref.2016.07.011}.
\bibitem[{Liu et~al.(2021)Liu, Zhang and Li}]{FJ-Liu-Zhang-Li-2021-cnSSC}
\bibinfo{author}{Liu, X.}, \bibinfo{author}{Zhang, X.}, \bibinfo{author}{Li,
  S.}, \bibinfo{year}{2021}.
\newblock \bibinfo{title}{{Measurement of China's macroeconomic resilience: A
  systemic risk perspective}}.
\newblock \bibinfo{journal}{Social Sciences in China (in Chinese)} ,
  \bibinfo{pages}{12--32+204.}\DOIprefix\doi{http://sscp.cssn.cn/zgshkx/zgshkx202101/202102/t20210201_5309025.html}.
\bibitem[{Lu et~al.(2018)Lu, Bessler and
  Leatham}]{FJ-Lu-Bessler-Leatham-2018-JIntFinancMarkInstMoney}
\bibinfo{author}{Lu, R.}, \bibinfo{author}{Bessler, D.A.},
  \bibinfo{author}{Leatham, D.J.}, \bibinfo{year}{2018}.
\newblock \bibinfo{title}{The transmission of liquidity shocks via {C}hina's
  segmented money market: Evidence from recent market events}.
\newblock \bibinfo{journal}{Journal of International Financial Markets
  Institutions \& Money} \bibinfo{volume}{57}, \bibinfo{pages}{110--126}.
\newblock \DOIprefix\doi{10.1016/j.intfin.2018.07.005}.
\bibitem[{Lyu et~al.(2021)Lyu, Wei, Hu and
  Yang}]{FJ-Lyu-Wei-Hu-Yang-2021-Energy}
\bibinfo{author}{Lyu, Y.}, \bibinfo{author}{Wei, Y.}, \bibinfo{author}{Hu, Y.},
  \bibinfo{author}{Yang, M.}, \bibinfo{year}{2021}.
\newblock \bibinfo{title}{Good volatility, bad volatility and economic
  uncertainty: Evidence from the crude oil futures market}.
\newblock \bibinfo{journal}{Energy} \bibinfo{volume}{222},
  \bibinfo{pages}{119924}.
\newblock \DOIprefix\doi{10.1016/j.energy.2021.119924}.
\bibitem[{Mamatzakis and
  Tsionas(2021)}]{FJ-Mamatzakis-Tsionas-2021-IntJFinancEcon}
\bibinfo{author}{Mamatzakis, E.C.}, \bibinfo{author}{Tsionas, M.G.},
  \bibinfo{year}{2021}.
\newblock \bibinfo{title}{A bayesian panel stochastic volatility measure of
  financial stability}.
\newblock \bibinfo{journal}{International Journal of Finance \& Economics}
  \bibinfo{volume}{26}, \bibinfo{pages}{5363--5384}.
\newblock \DOIprefix\doi{10.1002/ijfe.2070}.
\bibitem[{Mieg(2022)}]{FJ-Mieg-2022-RiskAnal}
\bibinfo{author}{Mieg, H.A.}, \bibinfo{year}{2022}.
\newblock \bibinfo{title}{Volatility as a transmitter of systemic risk: is
  there a structural risk in finance?}
\newblock \bibinfo{journal}{Risk Analysis} \bibinfo{volume}{42},
  \bibinfo{pages}{1952--1964}.
\newblock \DOIprefix\doi{10.1111/risa.13564}.
\bibitem[{Naifar(2024)}]{FJ-Naifar-2024-FinancResLett}
\bibinfo{author}{Naifar, N.}, \bibinfo{year}{2024}.
\newblock \bibinfo{title}{Climate policy uncertainty and comparative reactions
  across sustainable sectors: Resilience or vulnerability?}
\newblock \bibinfo{journal}{Finance Research Letters} \bibinfo{volume}{65},
  \bibinfo{pages}{105543}.
\newblock \DOIprefix\doi{10.1016/j.frl.2024.105543}.
\bibitem[{Nguyenhuu and \"{O}ersal(2024)}]{FJ-Nguyenhuu-Orsal-2024-WorldEcon}
\bibinfo{author}{Nguyenhuu, T.}, \bibinfo{author}{\"{O}ersal, D.K.},
  \bibinfo{year}{2024}.
\newblock \bibinfo{title}{Geopolitical risks and financial stress in emerging
  economies}.
\newblock \bibinfo{journal}{World Economy} \bibinfo{volume}{47},
  \bibinfo{pages}{217--237}.
\newblock \DOIprefix\doi{10.1111/twec.13529}.
\bibitem[{Obstfeld et~al.(2010)Obstfeld, Shambaugh and
  Taylor}]{FJ-Obstfeld-Shambaugh-Taylor-2010-AmEconJ-Macroecon}
\bibinfo{author}{Obstfeld, M.}, \bibinfo{author}{Shambaugh, J.C.},
  \bibinfo{author}{Taylor, A.M.}, \bibinfo{year}{2010}.
\newblock \bibinfo{title}{Financial stability, the trilemma, and international
  reserves}.
\newblock \bibinfo{journal}{American Economic Journal-Macroeconomics}
  \bibinfo{volume}{2}, \bibinfo{pages}{57--94}.
\newblock \DOIprefix\doi{10.1257/mac.2.2.57}.
\bibitem[{Pehrsson(2009)}]{FJ-Pehrsson-2009-IntMarketRev}
\bibinfo{author}{Pehrsson, A.}, \bibinfo{year}{2009}.
\newblock \bibinfo{title}{Marketing strategy antecedents of value adding by
  foreign subsidiaries}.
\newblock \bibinfo{journal}{International Marketing Review}
  \bibinfo{volume}{26}, \bibinfo{pages}{151--171}.
\newblock \DOIprefix\doi{10.1108/02651330910950402}.
\bibitem[{Phan et~al.(2021)Phan, Iyke, Sharma and
  Affandi}]{FJ-Phan-Iyke-Sharma-Affandi-2021-EconModel}
\bibinfo{author}{Phan, D.H.B.}, \bibinfo{author}{Iyke, B.N.},
  \bibinfo{author}{Sharma, S.S.}, \bibinfo{author}{Affandi, Y.},
  \bibinfo{year}{2021}.
\newblock \bibinfo{title}{Economic policy uncertainty and financial
  stability-{I}s there a relation?}
\newblock \bibinfo{journal}{Economic Modelling} \bibinfo{volume}{94},
  \bibinfo{pages}{1018--1029}.
\newblock \DOIprefix\doi{10.1016/j.econmod.2020.02.042}.
\bibitem[{Primiceri(2005)}]{FJ-Primiceri-2005-RevEconStud}
\bibinfo{author}{Primiceri, G.E.}, \bibinfo{year}{2005}.
\newblock \bibinfo{title}{Time varying structural vector autoregressions and
  monetary policy}.
\newblock \bibinfo{journal}{Review of Economic Studies} \bibinfo{volume}{72},
  \bibinfo{pages}{821--852}.
\newblock \DOIprefix\doi{10.1111/j.1467-937X.2005.00353.x}.
\bibitem[{Qiao et~al.(2024)Qiao, Ding, Han and
  Li}]{FJ-Qiao-Ding-Han-Li-2024-JIntMoneyFinan}
\bibinfo{author}{Qiao, T.}, \bibinfo{author}{Ding, W.}, \bibinfo{author}{Han,
  L.}, \bibinfo{author}{Li, D.}, \bibinfo{year}{2024}.
\newblock \bibinfo{title}{Rmb exchange rate volatility and the cross-section of
  {C}hinese a-share returns}.
\newblock \bibinfo{journal}{Journal of International Money and Finance}
  \bibinfo{volume}{142}, \bibinfo{pages}{103024}.
\newblock \DOIprefix\doi{10.1016/j.jimonfin.2024.103024}.
\bibitem[{Ren et~al.(2025)Ren, Li, Ji and
  Zhai}]{FJ-Ren-Li-Ji-Zhai-2025-ApplEconLett}
\bibinfo{author}{Ren, X.}, \bibinfo{author}{Li, Y.}, \bibinfo{author}{Ji, Q.},
  \bibinfo{author}{Zhai, P.}, \bibinfo{year}{2025}.
\newblock \bibinfo{title}{Climate policy uncertainty and the green bond market:
  fresh insights from the {QARDL} model}.
\newblock \bibinfo{journal}{Applied Economics Letters} \bibinfo{volume}{32},
  \bibinfo{pages}{464--469}.
\newblock \DOIprefix\doi{10.1080/13504851.2023.2275643}.
\bibitem[{Sahebjamnia et~al.(2015)Sahebjamnia, Torabi and
  Mansouri}]{FJ-Sahebjamnia-Torabi-Mansouri-2015-EurJOperRes}
\bibinfo{author}{Sahebjamnia, N.}, \bibinfo{author}{Torabi, S.A.},
  \bibinfo{author}{Mansouri, S.A.}, \bibinfo{year}{2015}.
\newblock \bibinfo{title}{Integrated business continuity and disaster recovery
  planning: Towards organizational resilience}.
\newblock \bibinfo{journal}{European Journal of Operational Research}
  \bibinfo{volume}{242}, \bibinfo{pages}{261--273}.
\newblock \DOIprefix\doi{10.1016/j.ejor.2014.09.055}.
\bibitem[{Sakyi-Nyarko et~al.(2022)Sakyi-Nyarko, Ahmad and
  Green}]{FJ-SakyiNyarko-Ahmad-Green-2022-JDevStud}
\bibinfo{author}{Sakyi-Nyarko, C.}, \bibinfo{author}{Ahmad, A.H.},
  \bibinfo{author}{Green, C.J.}, \bibinfo{year}{2022}.
\newblock \bibinfo{title}{The gender-differential effect of financial inclusion
  on household financial resilience}.
\newblock \bibinfo{journal}{Journal of Development Studies}
  \bibinfo{volume}{58}, \bibinfo{pages}{692--712}.
\newblock \DOIprefix\doi{10.1080/00220388.2021.2013467}.
\bibitem[{Seidl et~al.(2016)Seidl, Spies, Peterson, Stephens and
  Hicke}]{FJ-Seidl-Spies-Peterson-Stephens-Hicke-2016-JApplEcol}
\bibinfo{author}{Seidl, R.}, \bibinfo{author}{Spies, T.A.},
  \bibinfo{author}{Peterson, D.L.}, \bibinfo{author}{Stephens, S.L.},
  \bibinfo{author}{Hicke, J.A.}, \bibinfo{year}{2016}.
\newblock \bibinfo{title}{Searching for resilience: addressing the impacts of
  changing disturbance regimes on forest ecosystem services}.
\newblock \bibinfo{journal}{Journal of Applied Ecology} \bibinfo{volume}{53},
  \bibinfo{pages}{120--129}.
\newblock \DOIprefix\doi{10.1111/1365-2664.12511}.
\bibitem[{Sun et~al.(2021)Sun, Wu, Zeng and
  Peng}]{FJ-Sun-Wu-Zeng-Peng-2021-FinancResLett}
\bibinfo{author}{Sun, Y.}, \bibinfo{author}{Wu, M.}, \bibinfo{author}{Zeng,
  X.}, \bibinfo{author}{Peng, Z.}, \bibinfo{year}{2021}.
\newblock \bibinfo{title}{{The impact of COVID-19 on the Chinese stock market:
  Sentimental or substantial?}}
\newblock \bibinfo{journal}{Finance Research Letters} \bibinfo{volume}{38},
  \bibinfo{pages}{101838}.
\newblock \DOIprefix\doi{10.1016/j.frl.2020.101838}.
\bibitem[{Tang et~al.(2024)Tang, Liu and
  Yang}]{FJ-Tang-Liu-Yang-2024-Pac-BasinFinancJ}
\bibinfo{author}{Tang, C.}, \bibinfo{author}{Liu, X.}, \bibinfo{author}{Yang,
  G.}, \bibinfo{year}{2024}.
\newblock \bibinfo{title}{A study of financial market resilience in {C}hina -
  from a hot money shock perspective}.
\newblock \bibinfo{journal}{Pacific-Basin Finance Journal}
  \bibinfo{volume}{83}, \bibinfo{pages}{102256}.
\newblock \DOIprefix\doi{10.1016/j.pacfin.2024.102256}.
\bibitem[{Tang et~al.(2022)Tang, Liu and
  Zhou}]{FJ-Tang-Liu-Zhou-2022-JIntFinancMarkInstMoney}
\bibinfo{author}{Tang, C.}, \bibinfo{author}{Liu, X.}, \bibinfo{author}{Zhou,
  D.}, \bibinfo{year}{2022}.
\newblock \bibinfo{title}{Financial market resilience and financial
  development: a global perspective}.
\newblock \bibinfo{journal}{Journal of International Financial Markets
  Institutions \& Money} \bibinfo{volume}{80}, \bibinfo{pages}{101650}.
\newblock \DOIprefix\doi{10.1016/j.intfin.2022.101650}.
\bibitem[{Tiwari et~al.(2023)Tiwari, Suleman, Ullah and
  Shahbaz}]{FJ-Tiwari-Suleman-Ullah-Shahbaz-2023-IntJFinancEcon}
\bibinfo{author}{Tiwari, A.K.}, \bibinfo{author}{Suleman, M.T.},
  \bibinfo{author}{Ullah, S.}, \bibinfo{author}{Shahbaz, M.},
  \bibinfo{year}{2023}.
\newblock \bibinfo{title}{Analyzing the connectedness between crude oil and
  petroleum products: Evidence from {USA}}.
\newblock \bibinfo{journal}{International Journal of Finance \& Economics}
  \bibinfo{volume}{28}, \bibinfo{pages}{2278--2347}.
\newblock \DOIprefix\doi{10.1002/ijfe.2536}.
\bibitem[{Uddin et~al.(2021)Uddin, Chowdhury, Anderson and
  Chaudhuri}]{FJ-Uddin-Chowdhury-Anderson-Chaudhuri-2021-JBusRes}
\bibinfo{author}{Uddin, M.}, \bibinfo{author}{Chowdhury, A.},
  \bibinfo{author}{Anderson, K.}, \bibinfo{author}{Chaudhuri, K.},
  \bibinfo{year}{2021}.
\newblock \bibinfo{title}{The effect of {COVID}-19 pandemic on global stock
  market volatility: can economic strength help to manage the uncertainty?}
\newblock \bibinfo{journal}{Journal of Business Research}
  \bibinfo{volume}{128}, \bibinfo{pages}{31--44}.
\newblock \DOIprefix\doi{10.1016/j.jbusres.2021.01.061}.
\bibitem[{Wang et~al.(2016)Wang, Xie, Jiang and
  Stanley}]{FJ-Wang-Xie-Jiang-Stanley-2016-FinancResLett}
\bibinfo{author}{Wang, G.J.}, \bibinfo{author}{Xie, C.},
  \bibinfo{author}{Jiang, Z.Q.}, \bibinfo{author}{Stanley, H.E.},
  \bibinfo{year}{2016}.
\newblock \bibinfo{title}{Who are the net senders and recipients of volatility
  spillovers in {C}hina's financial markets?}
\newblock \bibinfo{journal}{Finance Research Letters} \bibinfo{volume}{18},
  \bibinfo{pages}{255--262}.
\newblock \DOIprefix\doi{10.1016/j.frl.2016.04.025}.
\bibitem[{Wang et~al.(2020)Wang, Li and
  He}]{FJ-Wang-Li-He-2020-ResIntBusFinanc}
\bibinfo{author}{Wang, Z.}, \bibinfo{author}{Li, Y.}, \bibinfo{author}{He, F.},
  \bibinfo{year}{2020}.
\newblock \bibinfo{title}{Asymmetric volatility spillovers between economic
  policy uncertainty and stock markets: Evidence from {C}hina}.
\newblock \bibinfo{journal}{Research in International Business and Finance}
  \bibinfo{volume}{53}, \bibinfo{pages}{101233}.
\newblock \DOIprefix\doi{10.1016/j.ribaf.2020.101233}.
\bibitem[{Wei et~al.(2024)Wei, Wang, Zhou, Shang and
  Ren}]{FJ-Wei-Wang-Zhou-Shang-Ren-2024-FinancResLett}
\bibinfo{author}{Wei, Y.}, \bibinfo{author}{Wang, Z.}, \bibinfo{author}{Zhou,
  X.}, \bibinfo{author}{Shang, Y.}, \bibinfo{author}{Ren, L.},
  \bibinfo{year}{2024}.
\newblock \bibinfo{title}{Are the leading indicators really leading? evidence
  from mixed-frequency spillover approach}.
\newblock \bibinfo{journal}{Finance Research Letters} \bibinfo{volume}{69},
  \bibinfo{pages}{106233}.
\newblock \DOIprefix\doi{10.1016/j.frl.2024.106233}.
\bibitem[{Wen et~al.(2021)Wen, Cao, Liu and
  Wang}]{FJ-Wen-Cao-Liu-Wang-2021-IntRevFinancAnal}
\bibinfo{author}{Wen, F.}, \bibinfo{author}{Cao, J.}, \bibinfo{author}{Liu,
  Z.}, \bibinfo{author}{Wang, X.}, \bibinfo{year}{2021}.
\newblock \bibinfo{title}{Dynamic volatility spillovers and investment
  strategies between the {C}hinese stock market and commodity markets}.
\newblock \bibinfo{journal}{International Review of Financial Analysis}
  \bibinfo{volume}{76}, \bibinfo{pages}{101772}.
\newblock \DOIprefix\doi{10.1016/j.irfa.2021.101772}.
\bibitem[{Wu et~al.(2023)Wu, Min and Wen}]{FJ-Wu-Min-Wen-2023-EurJFinanc}
\bibinfo{author}{Wu, B.}, \bibinfo{author}{Min, F.}, \bibinfo{author}{Wen, F.},
  \bibinfo{year}{2023}.
\newblock \bibinfo{title}{The stress contagion among financial markets and its
  determinants}.
\newblock \bibinfo{journal}{European Journal of Finance} \bibinfo{volume}{29},
  \bibinfo{pages}{1267--1302}.
\newblock \DOIprefix\doi{10.1080/1351847X.2022.2111222}.
\bibitem[{Wu et~al.(2024)Wu, Li and Su}]{FJ-Wu-Li-Su-2024-EconAnalPolicy}
\bibinfo{author}{Wu, J.}, \bibinfo{author}{Li, J.P.}, \bibinfo{author}{Su,
  C.W.}, \bibinfo{year}{2024}.
\newblock \bibinfo{title}{{Can green bond hedges climate policy uncertainty in
  the United States: new insights from novel time-varying causality and
  quantile-on-quantile methods?}}
\newblock \bibinfo{journal}{Economic Analysis and Policy} \bibinfo{volume}{82},
  \bibinfo{pages}{1158--1176}.
\newblock \DOIprefix\doi{10.1016/j.eap.2024.04.015}.
\bibitem[{Xu and Liu(2024)}]{FJ-Xu-Liu-2024-EmergMarkRev}
\bibinfo{author}{Xu, D.}, \bibinfo{author}{Liu, Y.}, \bibinfo{year}{2024}.
\newblock \bibinfo{title}{{How does technological progress affect provincial
  financial resilience? Evidence at the provincial level in China}}.
\newblock \bibinfo{journal}{Emerging Markets Review} \bibinfo{volume}{60},
  \bibinfo{pages}{101137}.
\newblock \DOIprefix\doi{10.1016/j.ememar.2024.101137}.
\bibitem[{Yang et~al.(2024)Yang, Geng and
  Liang}]{FJ-Yang-Geng-Liang-2024-EnergyEcon}
\bibinfo{author}{Yang, J.}, \bibinfo{author}{Geng, J.B.},
  \bibinfo{author}{Liang, Z.}, \bibinfo{year}{2024}.
\newblock \bibinfo{title}{Time-varying effects of structural oil price shocks
  on financial market uncertainty}.
\newblock \bibinfo{journal}{Energy Economics} \bibinfo{volume}{139},
  \bibinfo{pages}{107910}.
\newblock \DOIprefix\doi{10.1016/j.eneco.2024.107910}.
\bibitem[{Yao et~al.(2025)Yao, Maimaitijiang, Li and
  Le}]{FJ-Yao-Maimaitijiang-Li-Le-2025-JCommodMark}
\bibinfo{author}{Yao, X.}, \bibinfo{author}{Maimaitijiang, S.},
  \bibinfo{author}{Li, J.}, \bibinfo{author}{Le, W.}, \bibinfo{year}{2025}.
\newblock \bibinfo{title}{How financial markets respond to climate policy
  uncertainty: a dynamic resilience analysis}.
\newblock \bibinfo{journal}{Journal of Commodity Markets} \bibinfo{volume}{39},
  \bibinfo{pages}{100490}.
\newblock \DOIprefix\doi{10.1016/j.jcomm.2025.100490}.
\bibitem[{You et~al.(2017)You, Guo, Zhu and
  Tang}]{FJ-You-Guo-Zhu-Tang-2017-EnergyEcon}
\bibinfo{author}{You, W.}, \bibinfo{author}{Guo, Y.}, \bibinfo{author}{Zhu,
  H.}, \bibinfo{author}{Tang, Y.}, \bibinfo{year}{2017}.
\newblock \bibinfo{title}{Oil price shocks, economic policy uncertainty and
  industry stock returns in {C}hina: {A}symmetric effects with quantile
  regression}.
\newblock \bibinfo{journal}{Energy Economics} \bibinfo{volume}{68},
  \bibinfo{pages}{1--18}.
\newblock \DOIprefix\doi{10.1016/j.eneco.2017.09.007}.
\bibitem[{Zhang et~al.(2025)Zhang, Shi, Zhao and
  Yang}]{FJ-Zhang-Shi-Zhao-Yang-2025-RiskManag}
\bibinfo{author}{Zhang, Q.}, \bibinfo{author}{Shi, Y.}, \bibinfo{author}{Zhao,
  X.}, \bibinfo{author}{Yang, J.}, \bibinfo{year}{2025}.
\newblock \bibinfo{title}{Climate vulnerability, macroprudential policy, and
  financial risk}.
\newblock \bibinfo{journal}{Risk Management} \bibinfo{volume}{27},
  \bibinfo{pages}{25}.
\newblock \DOIprefix\doi{10.1057/s41283-025-00179-3}.
\bibitem[{Zhao et~al.(2020)Zhao, Zhang and
  Liu}]{FJ-Zhao-Zhang-Liu-2020-EmergMarkFinancTrade}
\bibinfo{author}{Zhao, X.}, \bibinfo{author}{Zhang, W.G.},
  \bibinfo{author}{Liu, Y.J.}, \bibinfo{year}{2020}.
\newblock \bibinfo{title}{Volatility spillovers and risk contagion paths with
  capital flows across multiple financial markets in {C}hina}.
\newblock \bibinfo{journal}{Emerging Markets Finance and Trade}
  \bibinfo{volume}{56}, \bibinfo{pages}{731--749}.
\newblock \DOIprefix\doi{10.1080/1540496X.2018.1472080}.
\bibitem[{Zhu et~al.(2025)Zhu, Xia, Li and
  Chen}]{FJ-Zhu-Xia-Li-Chen-2025-FinancResLett}
\bibinfo{author}{Zhu, S.}, \bibinfo{author}{Xia, Y.}, \bibinfo{author}{Li, Q.},
  \bibinfo{author}{Chen, Y.}, \bibinfo{year}{2025}.
\newblock \bibinfo{title}{Global geopolitical risk and financial stability:
  Evidence from {C}hina}.
\newblock \bibinfo{journal}{Finance Research Letters} \bibinfo{volume}{72},
  \bibinfo{pages}{106501}.
\newblock \DOIprefix\doi{10.1016/j.frl.2024.106501}.
\bibitem[{Zou(2024)}]{FJ-Zou-2024-EconModel}
\bibinfo{author}{Zou, Y.}, \bibinfo{year}{2024}.
\newblock \bibinfo{title}{The impact of fiscal stimulus on employment: Evidence
  from {C}hina's four-trillion rmb package}.
\newblock \bibinfo{journal}{Economic Modelling} \bibinfo{volume}{131},
  \bibinfo{pages}{106598}.
\newblock \DOIprefix\doi{10.1016/j.econmod.2023.106598}.

\end{thebibliography}

\clearpage

\end{document}